\documentclass[english,twocolumn,prl,preprintnumbers,amsmath,amssymb,superscriptaddress]{revtex4-1}
\usepackage[normalem]{ulem}
\usepackage[utf8]{inputenc}
\usepackage[english]{babel}
\usepackage{xcolor}
\usepackage{epstopdf}
\usepackage{verbatim}
\usepackage{graphicx}
\usepackage{amssymb}
\usepackage{babel}
\makeatother

\newcommand{\bra}[1]{\left\langle {#1} \right|}
\newcommand{\ket}[1]{\left|  #1 \right\rangle}

\newcommand{\RN}[1]{%
	\textup{\uppercase\expandafter{\romannumeral#1}}
}

\begin{document}
	

	\title{Observation of quantum phase synchronization in spin-1 atoms}
	
	\author{Arif Warsi Laskar}
	\affiliation{Department of Physics, Indian Institute of Technology - Kanpur, UP-208016, India}
	\author{Pratik Adhikary}
	\affiliation{Department of Physics, Indian Institute of Technology - Kanpur, UP-208016, India}
	\author{Suprodip Mondal}
	\affiliation{Department of Physics, Indian Institute of Technology - Kanpur, UP-208016, India}
	\author{Parag Katiyar}
	\affiliation{Department of Physics, Indian Institute of Technology - Kanpur, UP-208016, India}
	\author{Sai Vinjanampathy}
	\affiliation{Department of Physics, Indian Institute of Technology-Bombay, Powai, Mumbai 400076, India}
	\affiliation{Centre for Quantum Technologies, National University of Singapore, 3 Science Drive 2, 117543 Singapore, Singapore}
	\author{Saikat Ghosh}
	\email{gsaikat@iitk.ac.in}
	\affiliation{Department of Physics, Indian Institute of Technology - Kanpur, UP-208016, India}

\begin{abstract}
With growing interest in quantum technologies, possibilities of synchronizing quantum systems has garnered significant recent attention. In experiments with dilute ensemble of laser cooled spin-1 $^{87}{Rb}$ atoms, we observe phase difference of spin coherences to synchronize with phases of external classical fields. An initial limit-cycle state of a spin-1 atom localizes in phase space due to dark-state-polaritons generated by classical two-photon tone fields. In particular, when the two couplings fields are out of phase, the limit-cycle state synchronizes only with two artificially engineered, anisotropic decay rates. Furthermore, we observe a blockade of synchronization due to quantum interference and emergence of Arnold tongue-like features. Such anisotropic decay induced synchronization of spin-1 systems with no classical analogue can provide insights in open quantum systems and find applications in synchronized quantum networks.
 
\end{abstract}

	\date{\today}

\maketitle

Spontaneous synchronization is abundant in nature, ranging from synchronized fire-flies to neuronal activities~\cite{pikovskij,Strogatz08}. Such synchronous dynamics, being stable to external perturbations, have also found a range of applications including satellites~\cite{Lewandowski11}, electrical grids~\cite{Witthaut12}, clocks~\cite{Nathan16} and wind turbines~\cite{Gonzalez02}. Recently, synchronization in quantum domain has emerged as a field for understanding correlations~\cite{Kimble08,Zambrini13,Fazio13,Bruder14,Shenshen14,Bruder16,Marquardt17,Bruder17,Weiss17,Roulet18,Leong-Chuan18,Roulet19,Bruder18} and for applications in quantum networks~\cite{Duan01,Kimble08,Sadeghpour13}. Early proposals focused on open quantum systems whose mean-field theories exhibited synchronization~\cite{Sadeghpour13,Bruder14,Witthaut17,Brandes15}. Such models were extended deep in the quantum regime and compared to finite dimensional systems that have no classical analogues~\cite{Bruder16,Roulet18,Zambrini19}. In these systems, suitably chosen angles in phase space were found to be entrained to the phase of an external signal. Despite such proposals, observation of synchronization deep in the quantum regime has remained elusive.

\begin{figure*}
	\centering
	\includegraphics[scale=0.43]{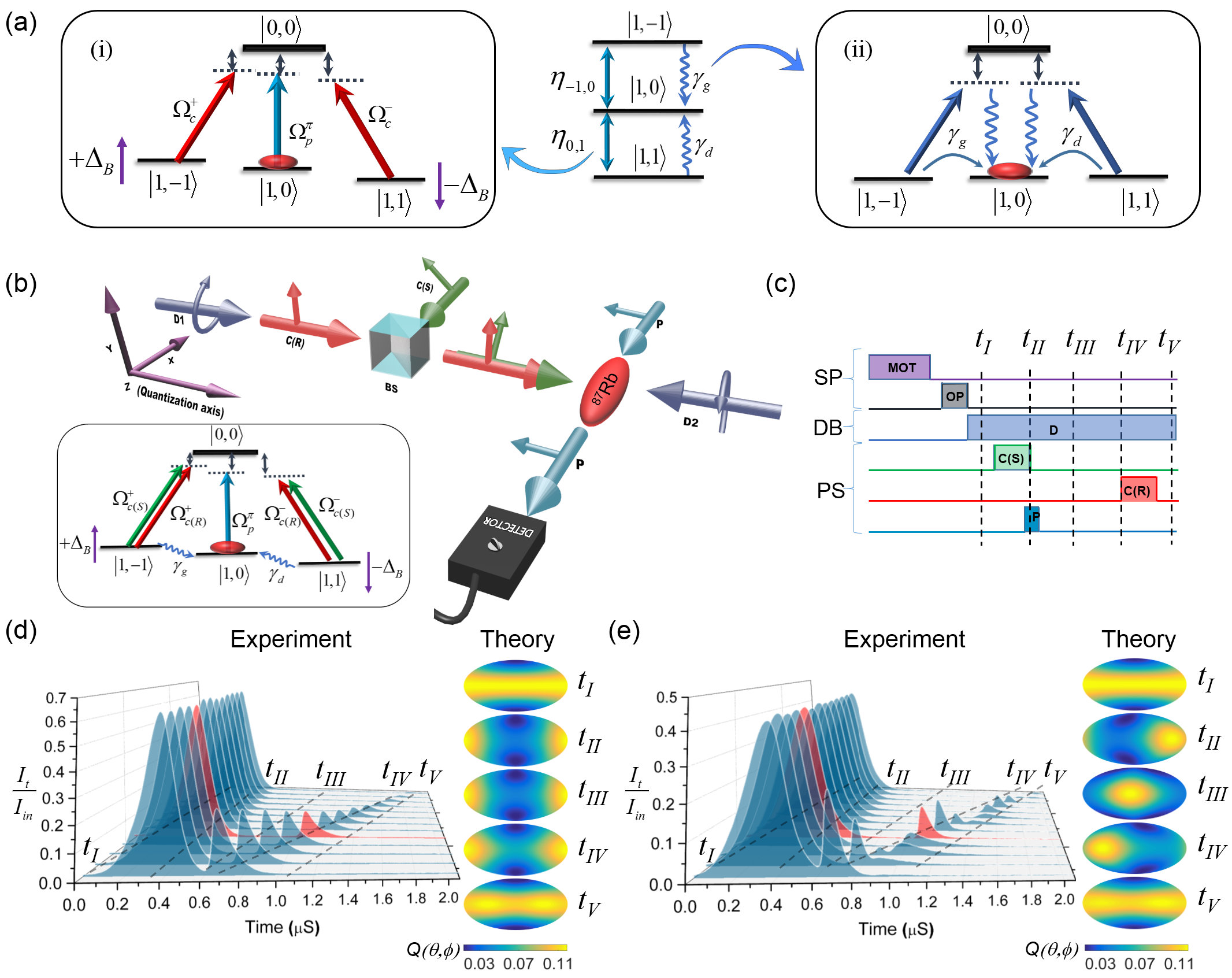}
	\caption{(a) Energy level diagram of a spin-1 atom with coherent couplings $\eta_{-1,0}$ and $\eta_{0,1}$ along with \textit{incoherent} decay rates $\gamma_g$ and $\gamma_d$. (a.i) Coherent couplings are engineered using control ($\Omega_c^{\pm}$) and probe fields ($\Omega_p^{\pi}$). Here $\Delta_B$ is the ground state energy shift due to a magnetic field along the quantization axis. (a.ii) Decay rates $\gamma_g$ and $\gamma_d$ are engineered with two fields coupling states $\ket{1,-1}$ and $\ket{1,1}$ to the excited state $\ket{0,0}$. (b) Experimental setup, showing propagation directions of storing (C(S), green) and retrieving (C(R), red) control fields, probe field (P, blue) and decay beams (D1 and D2, in violet, at small angles to control fields). Here BS: beam splitter. (c) Experimental timing sequence with intervals for state preparation (SP), decay beams (DB) and control-probe (DSP) fields. Here MOT: magneto optical trap; OP: optical pumping; D: decay beams. (d) Typical probe field time traces with varying storage times and with $\phi_c=\phi_d=0$. Reconstructed Husimi-Q function from numerical simulations (corresponding to red time trace) are plotted at different times (with the horizontal and vertical axes corresponding to $\phi$ and $\theta$, respectively and $0\leq\theta\leq\pi$ and $-\pi\leq\phi\leq\pi$). (e) In presence of magnetic field (with dynamic phase $\phi_d/\pi=3.22$, the tone phase set to zero), the retrieved intensity oscillates with changing storage time. Corresponding Husimi-Q function gets localized and entrained with the dynamic phase, precessing in equatorial plane. $I_{in}$ and $I_t$ are input and transmitted probe intensities, respectively.}
	\label{fig:interferometer}
\end{figure*}

\begin{figure*}
\centering
	\includegraphics[scale=0.31]{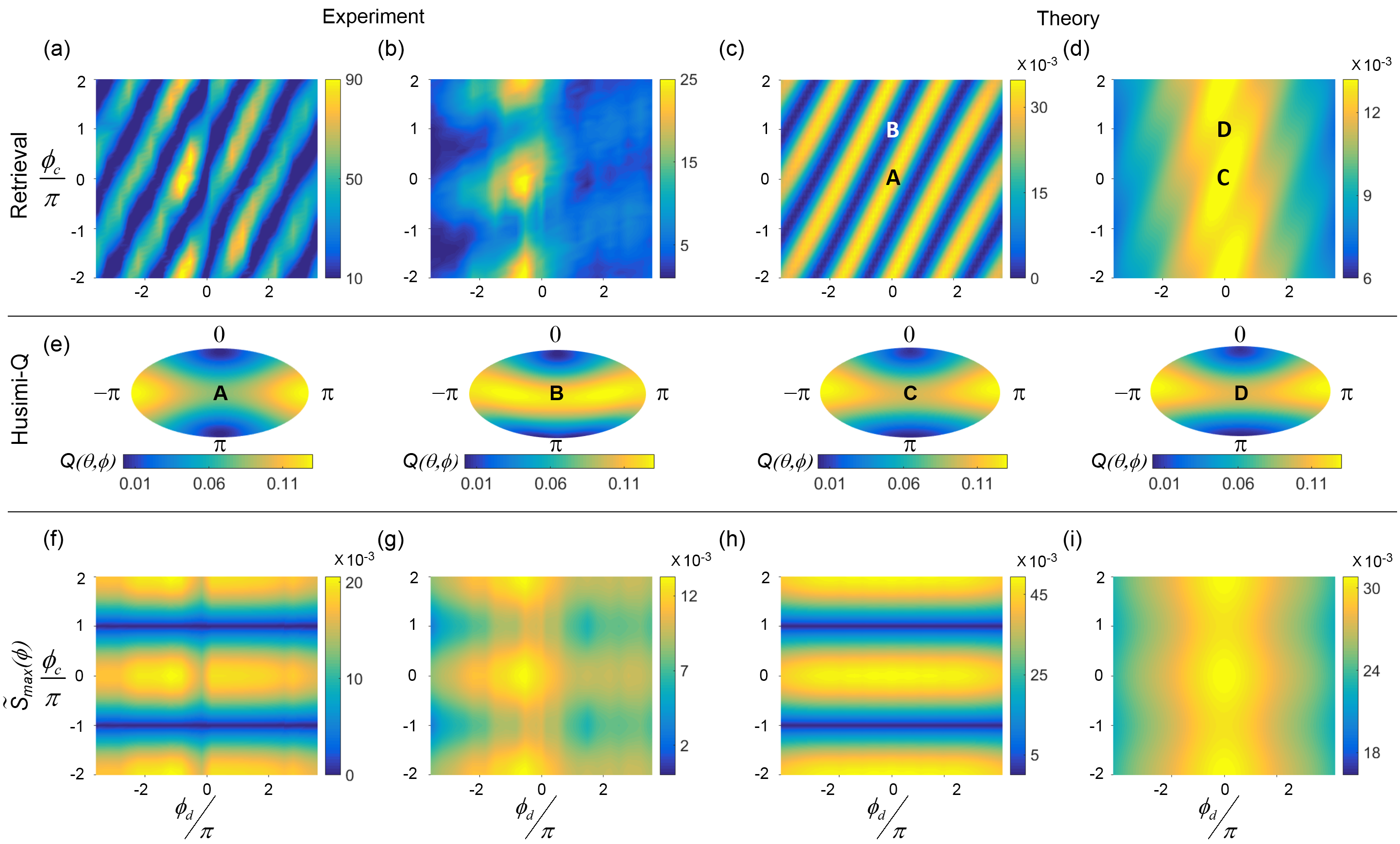}
	\caption{(a,b) Plots of retrieved intensity with dynamic ($\phi_d$) and tone ($\phi_c$) phases and with decay rate ratios: (a) $\gamma_d/\gamma_g=$ 1.00 and (b) 11.90 . $\phi_d$ is varied with magnetic field while keeping the storage time ($\tau$ = 600 ns) fixed. (c,d) Numerically simulated retrieved intensity for decay rate ratios: $\gamma_d/\gamma_g=$ 1.00 (c) and 11.00 (d). (e) Husimi-Q plots for the regions A, B, C, and D of (c) and (d). (f-i) Maximum of synchronization function ($\tilde{S}_{max}(\phi)$) calculated from experimental and simulated plots corresponding to (a),(b),(c) and (d). Here experimental parameters $\Omega^\pi_p$, $\Omega^{\text{lin}}_{c(S)}$, $\Omega^{\text{lin}}_{c(R)}$ and $\gamma_g$ are set to 0.64$\gamma$, 1.02$\gamma$, 1.44$\gamma$ and 107 kHz, respectively, and the simulation parameters are as tabulated in SI. Color bars for experimental plots are in units of $\mu$W/cm$^2$ and for theory plots, in units of $2|\Omega_R|^2/\gamma^2$, where $|\Omega_R|$ and $\gamma$ correspond to retrieved field and excited state decay, respectively. 
	}
	\label{fig:synchronization}
\end{figure*}
\begin{figure}
	\centering
	\includegraphics[scale=0.5]{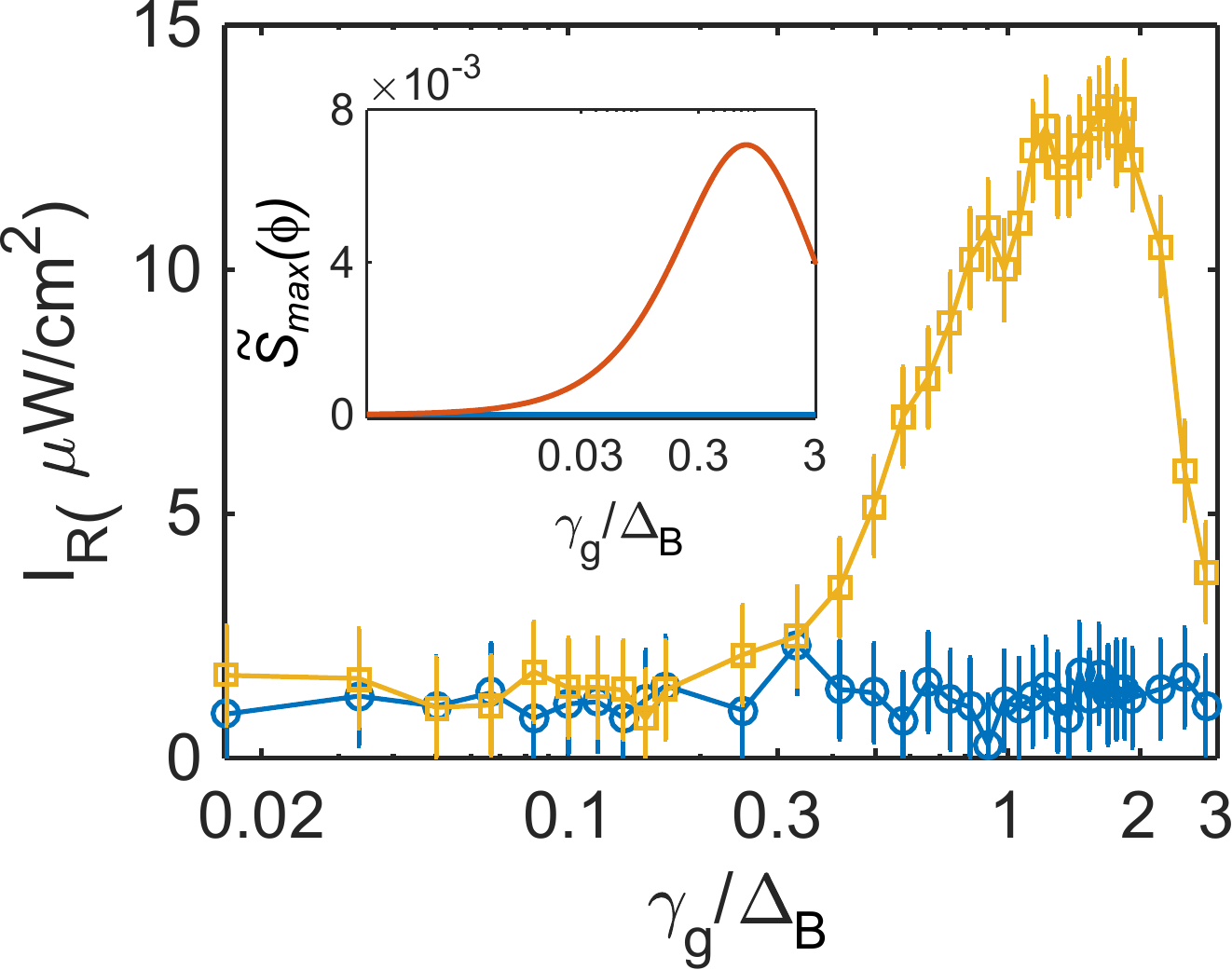}
	\caption{Retrieved intensity at a region of destructive interference is plotted as a function of increasing $\gamma_g$, for a fixed $\Delta_B$. Blue and yellow are for $\gamma_d/\gamma_g=$  1.00 and 11.90, respectively. Inset shows $\tilde{S}_{max}(\phi)$, as extracted from experimental data. Blue and red correspond to Figs. 2(a) and 2(b), respectively. Here the experimental parameters $\Omega^\pi_p$, $\Omega^{\text{lin}}_{c(S)}$ and $\Omega^{\text{lin}}_{c(R)}$ are set as in Fig.~\ref{fig:synchronization} with $\phi_c = 140^{\circ}$ and $\Delta_B=67$ kHz.}
	\label{fig:Blockade}
\end{figure}

Recently, it was pointed out that spin-1 is the smallest quantum system with a limit-cycle in phase space that can synchronize to external tone phases ~\cite{Roulet18,Roulet19}. Furthermore,  Roulet \textit{et. al.} predicted that weak classical tones (represented as two coherent couplings $\eta_{-1,0}, ~\eta_{0,1}$, Fig.~\ref{fig:interferometer}(a)) and anisotropic internal decay rates $\gamma_g$ and $\gamma_d$ can localize and synchronize the limit-cycle state, for all tone phases.

\par Here we report first observation of quantum synchronization with spin-1 systems, realized in a dilute ensemble of approximately a million laser-cooled $^{87}Rb$ atoms in $\ket{F=1}$ hyperfine ground-state manifold. Atoms are initialized to a limit-cycle in phase space, corresponding to the state $\ket{F=1, m_F=0}$  (Fig.~\ref{fig:interferometer}(a) and \ref{fig:interferometer}(b)). Synchronization is initiated with two circularly polarized \textit{control} fields ($\Omega_{c(S)}^{\pm}$) along with a $\pi$-polarized \textit{probe} $\Omega_p^{\pi}$. These fields induce coherent two-photon couplings between the spin states $\ket{F=1, m_F=\pm 1}$ and $\ket{F=1, m_F=0}$ (Fig.~\ref{fig:interferometer}(a), inset(i))~\cite{arif18}, corresponding to the weak tones $\eta_{-1,0}, ~\eta_{0,1}$~\cite{Roulet18,Roulet19}. When the control fields are adiabatically switched off, the probe gets stored as two dark state polaritons (DSPs) in atomic coherences $\rho_{-1,0}$ and $\rho_{0,1}$~\cite{Lukin00,Hau01,Lukin01,Lukin02,Martin08,Xiao11,Kuzmich13}. In the dark, the DSPs evolve in time, acquiring relative dynamic phase. From the retrieved DSPs as optical fields, we estimate the coherences and reconstruct a measure of synchronization. In particular, we observe a non-zero synchronization, for all tone phases, only when the two decay rates $\gamma_g$ and $\gamma_d$ are anisotropic (Fig.~\ref{fig:interferometer}(a), inset(ii)). We further observe a synchronization blockade due to destructive interference and emergence of Arnold-tongue like features with increasing drive: these have been predicted as quintessential signatures of quantum synchronization~\cite{Roulet18,Roulet19}.

\par Fig.~\ref{fig:interferometer}(d) and \ref{fig:interferometer}(e) show typical experimental time traces for probe pulses, with progressively increasing time of storage, with and without applied magnetic field, respectively. After cooling in a magneto-optic trap, the atoms are optically pumped in the ground state $\ket{F=1, m_F=0}$. A linearly polarized control, comprising two circularly polarized fields, is adiabatically switched on (at time instance $t_{I}$ in Fig.~\ref{fig:interferometer}(c)) and off (time $t_{II}$), after 1.6 $\mu$s, such that a probe pulse gets partially stored (see Supplementary Information (SI): S.V.(a)). The corresponding coherences evolve and interfere in the dark due to applied magnetic field (interval $t_{III}$, see SI: S.III.). We observe oscillations in the retrieved pulse as the storage time ($\tau=t_{IV}-t_{II}$) is varied (Fig.~\ref{fig:interferometer}(e)).   

Numerically simulated time traces, in close agreement with observations (see SI: S.IV.), are used to reconstruct the underlying spin-1 atomic state (corresponding to $\ket{F=1}$ manifold). In particular, a state $\hat{\rho}$ is visualized using Husimi-Q function, defined as~\cite{Gilmore,HUSIMI40,Monica15} 
\begin{eqnarray}
Q(\theta,\phi)=\frac{3}{4\pi}\bra{\theta,\phi}\hat{\rho}\ket{\theta,\phi}. \nonumber
\end{eqnarray}
Here 
$\ket{\theta,\phi}=\\
\text{cos}^{2}\frac{\theta}{2}\left(\ket{-1}+\sqrt{2}e^{i\phi}\text{tan}\left(\frac{\theta}{2}\right)\ket{0}+e^{i2\phi}\text{tan}^{2}\left(\frac{\theta}{2}\right)\ket{1}\right)$ is a spin coherent state~\cite{Radcliffe71}, parametrized by the angles $\theta$ and $\phi$. 

From Husimi-Q functions, plotted using Hammer projection (shown in red, Fig.~\ref{fig:interferometer}(d) and \ref{fig:interferometer}(e)), we note the initial state $\ket{F=1, m_F=0}$ corresponds to a limit-cycle (see SI: S.II.)~\cite{Roulet18,Roulet19}. Between time instances $t_{II}$ and $t_{IV}$,  the two in-phase ($\phi_c=0$) control and the probe fields result in a localized state in phase space (Fig.~\ref{fig:interferometer}(d) and \ref{fig:interferometer}(e) ``Theory" plots in right panels). In particular, with a magnetic field ($B_z$) along the quantization axis (Fig.~\ref{fig:interferometer}(b)), the state gets entrained and precesses in the equatorial plane, resulting in an accumulated dynamic phase $\phi_d$ = 2$\Delta_B\tau$ over a total storage time $\tau$ (Fig.~\ref{fig:interferometer}(e)). Here $\Delta_B=\mu_{B}B_z/2\hbar$ is the ground state shifts due to $B_z$ (Fig.~\ref{fig:interferometer}(a)) and $\mu_{B}$ is the Bohr magneton.

\begin{figure*}
	\centering
	\includegraphics[scale=0.5]{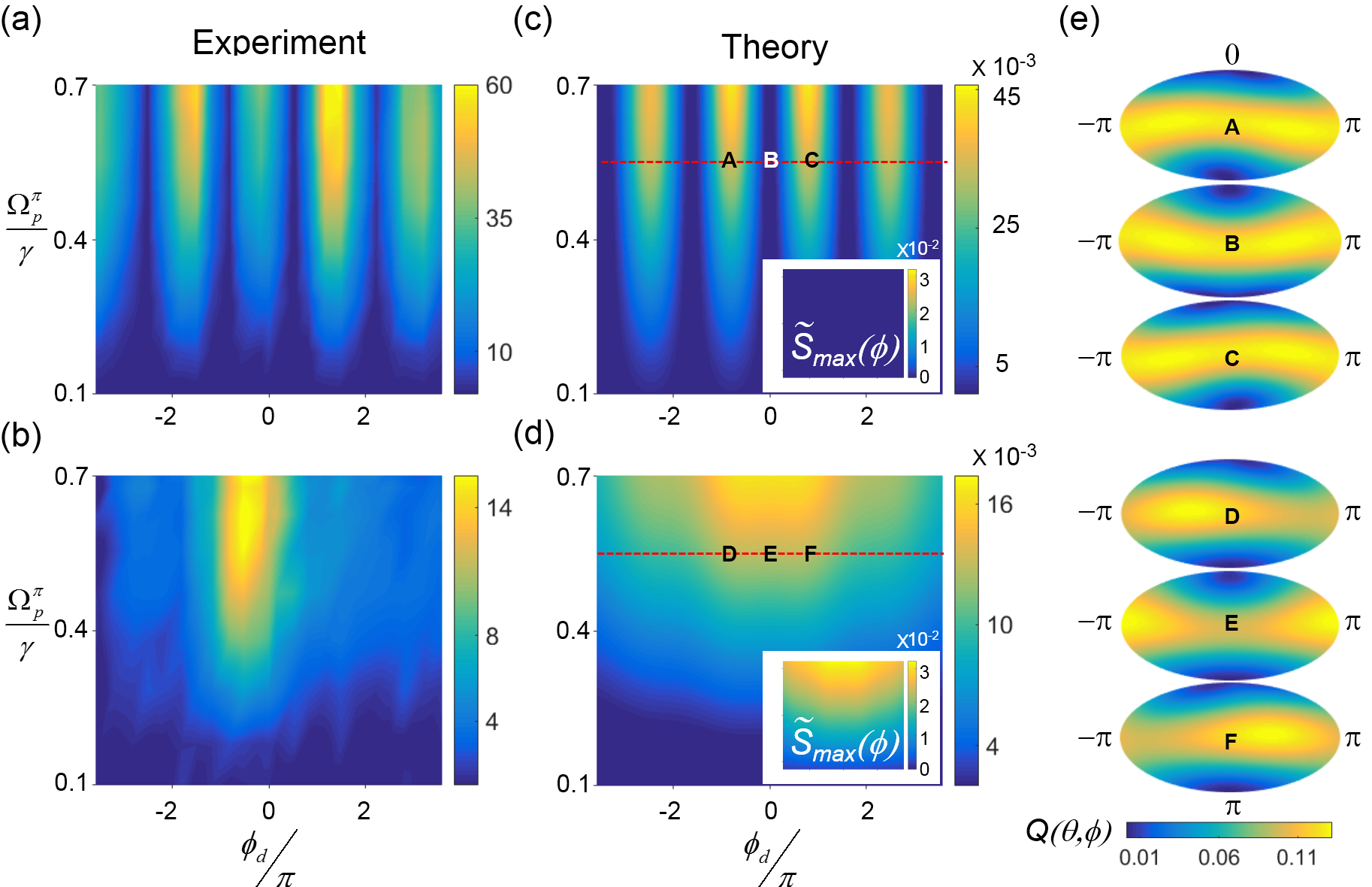}
	\caption{(a,b) Retrieved intensity, plotted with increasing probe field strength ($\Omega^\pi_p$) and dynamic phase ($\phi_d$) (storage time $\tau$ = 600 ns), with $\gamma_d/\gamma_g=$ 1 and 11.90 for (a) and (b), respectively. (c,d) Simulated plots for $\gamma_d/\gamma_g=$ 1.00 (c), and 11.00 (d). Insets of (c) and (d) are the corresponding $\tilde{S}_{max}(\phi)$. (e) Husimi-Q plots for regions A,B,C and D,E,F of (c) and (d), respectively. Here experimental parameters $\Omega^{\text{lin}}_{c(S)}$, $\Omega^{\text{lin}}_{c(R)}$, $\gamma_g$ are set as in Fig.~\ref{fig:synchronization} and $\phi_c$ is set to 140$^\circ$. Simulation parameters are as tabulated in SI. Colour bars for plots (a,b) are as in Fig. 2.}
	\label{fig:Arnold}
\end{figure*}

Fig.~\ref{fig:synchronization}(a) (\ref{fig:synchronization}(c)) shows typical experimental (simulation) plots of the retrieved intensity, $I_R(\tau)$ at $\tau=600$ ns, with changing tone ($\phi_c$) and the dynamic phase ($\phi_d$) (see SI: S.III.). Tone phase is varied by changing the phase difference between the stored and retrieving control fields, and dynamic phase is scanned with a magnetic field (Fig.~\ref{fig:interferometer}(b)). In particular, at regions \textbf{A} (corresponding to $\phi_c=0$) and \textbf{B} ($\phi_c=\pi$), the states in phase space appear dramatically different (Fig.~\ref{fig:synchronization}(c)). While the state gets localized at \textbf{A}, it remains a delocalized limit-cycle at \textbf{B} (Fig.~\ref{fig:synchronization}(e), \textbf{A} and \textbf{B}, respectively). However, when the artificially engineered decay rates (see SI: S.V.(d)) are made anisotropic ($\gamma_d/\gamma_g=$ 11.00, Fig.~\ref{fig:synchronization}(b) and \ref{fig:synchronization}(d) for experiment and simulations, respectively), the state localizes for all tone phases including $\phi_c=0$ and $\phi_c=\pi$ (Fig.~\ref{fig:synchronization}(e), \textbf{C} and \textbf{D}) respectively). 

These localized states can be quantified with a synchronization function, defined as~\cite{Roulet19} (see SI: S.II.):
\begin{align}\label{syncfn}
&S(\phi)=\int_{0}^{\pi}Q(\theta,\phi)\sin\theta d\theta-\frac{1}{2\pi},\nonumber \\ 
&\simeq \frac{3}{8\sqrt{2}}[{|\rho_{-1,0}|\cos(\Delta_B\tau+\phi)+|\rho_{0,1}|\cos(\Delta_B\tau+\phi+\phi_c)}]\nonumber
\end{align}
(with squeezing tone, $\rho_{-1,1}\simeq 0$). For a limit cycle state, $S(\phi)$ remains identically zero. On the contrary, any non-zero $S(\phi)$ indicate localized, synchronized states in phase space. From the visibility of the interference fringes with $\phi_c$ (Fig. 2(a),2(b),2(c) and 2(d)), we estimate the coherences $|\rho_{0,1}|$, $|\rho_{-1,0}|$ and their relative phases, thereby reconstructing the measure from experimental data (Fig.~\ref{fig:synchronization}(f) and \ref{fig:synchronization}(g)) data and compare it with simulations (Fig.~\ref{fig:synchronization}(h) and \ref{fig:synchronization}(i); see SI: S.II.). Our estimated synchronization function $\tilde{S}(\phi)$, is related to the measure as $S(\phi) =\kappa \tilde{S}(\phi)$, with $\kappa$ depending on normalization, retrieved field strength and optical depth.

We observe maximum of $\tilde{S}(\phi)$ i.e. $\tilde{S}_{max}(\phi)$ to remain non-zero when the two tones are in phase ($\phi_c=0$, Fig.~\ref{fig:synchronization}(f) and \ref{fig:synchronization}(h)). On the contrary, when the two tones are out-of phase ($\phi_c=\pi$), $\tilde{S}_{max}(\phi)$ sharply falls to zero. However, for anisotropic decay rates (Fig.~\ref{fig:synchronization}(g) and \ref{fig:synchronization}(i)), $\tilde{S}_{max}(\phi)$ is non-zero for all $\phi_c$. 

For out-of-phase tones, there is a rich interplay between incoherent and coherent dynamics ~\cite{arif16,arif18}. When both decay rates are smaller than $\Delta_B$, the coherences destructively interfere and the state remains a limit cycle. However, as the decay rates become comparable to $\Delta_B$, the coherences interfere only partially, resulting in a localized state. In Fig.~\ref{fig:Blockade}, over an extended region ($\gamma_g<\Delta_B$), destructive interference (with  $\phi_c= 140^\circ$) causes a blockade of synchronization. Moreover, as the decay rates are made anisotropic ($\gamma_d>\gamma_g$), the estimated $\tilde{S}_{max}(\phi)$ at a fixed $\phi_c$ becomes non-zero along with a finite retrieval. When both decay rates are large ($\gamma_d,\gamma_g>\Delta_B$), there is overall decrease in retrieval due to decoherence. Such \textit{blockade} of synchronization due to quantum interference and re-emergence of entrained states are typical quantum signatures~\cite{Bruder17,Roulet19}.

The state synchronizes over a range of dynamic phase ($\phi_d$) and field strengths. Such dependence, leading to Arnold-tongue like features, have been studied as typical signatures in classical and quantum synchronization~\cite{pikovskij}. Here, for out-of-phase tone and with increasing probe field ($\Omega_p^{\pi}$) we observe fringes with varying magnetic field (Fig.~\ref{fig:Arnold}(a) and \ref{fig:Arnold}(c)). For equal decays, the states remain delocalized (typical regions \textbf{A}, \textbf{B} and \textbf{C}, Fig.~\ref{fig:Arnold}(e)) with $\tilde{S}_{max}(\phi)\sim 0$ for all dynamic phases (estimated from simulations, Fig.~\ref{fig:Arnold}(c), inset). Furthermore, for anisotropic rates ($\gamma_d/\gamma_g=11.0$, Fig.~\ref{fig:Arnold}(b) and \ref{fig:Arnold}(d)), the fringes merge to a single maxima, broadening into a Arnold-tongue like shape with increasing $\Omega_p^{\pi}$. Since $\phi_c$ is kept constant,  $\tilde{S}_{max}(\phi)$ could not be evaluated from fringe visibility with tone phase. Nevertheless, when evaluated numerically, Arnold-tongue for $\tilde{S}_{max}(\phi)$ emerges from a null background (Fig.~\ref{fig:Arnold}(d), inset) along with localized states (regions \textbf{D}, \textbf{E} and \textbf{F}, Fig.~\ref{fig:Arnold}(d)) and entrained with $\phi_d$ (\textbf{D}, \textbf{E} and \textbf{F}, Fig.~\ref{fig:Arnold}(e)).

To conclude, here we report first observation of synchronization in the smallest quantum system. Synchronization is achieved using two primary resources: quantum coherence and engineered decay rates. In particular, for anisotropic decay rates, limit-cycle states synchronize for all tone phases. Furthermore, we observe two typical quantum signatures: a synchronization blockade due to quantum interference and emergence of Arnold tongue in  $\tilde{S}_{max}(\phi)$. The experimentally observed Arnold tongue is narrower, which can bear signatures of superradiance~\cite{Dicke54}. A search for synchronization in multiple such spin systems, towards synchronized quantum memories can lead to direct applications in quantum networks.  

We thank S. Chakraborty, S. Das, A. Gupta, M.Hajdu\v{s}ek, U. Rapol and H. Wanare for insightful discussions and comments. SV, PA and SM acknowledge support from DST-SERB Early Career Research Award (ECR/2018/000957), UGC and CSIR, respectively. This work was supported under SERB-DST grant no: SERB/PHY/2015404. 


%

\clearpage

\begin{widetext}

\begin{center}
\textbf{\large Supplementary information:}
\end{center}

Here, we provide details of numerical simulations used to reconstruct the spin-1 states and discuss construction of the Husimi-Q function for visualization of the state in phase space. Furthermore, we provide an analytic model describing interferometry of dark-state-polaritons (DSPs), that is used to interpret signatures of quantum synchronization. We also document some experimental details pertaining to the experiment and measurements presented here.

\bigskip
\noindent
\textbf{S.I. Methods}

\noindent
Experiment:

\textit{Atomic system:} 
$^{87}Rb$ atoms are laser-cooled and trapped in a cigar-shaped cloud at a temperature of $\sim$121$\mu$K. After turning off all cooling and trapping fields, the atoms are pumped to the state $\ket{F=1,m_F=0}$ using a pair of optical fields by first placing the atoms in $F=1$ ground state manifold and then spin polarizing them to the particular state. 

\textit{Coherent couplings (tones):} Two kinds of couplings are engineered for the experiment-\textit{coherent} two-photon couplings and \textit{incoherent} decay rates between states. 
For \textit{coherent} couplings, a linearly polarized laser ($\Omega_{c(S)}^{\text{lin}}$), composed of two circularly polarized control fields ($\Omega_{c(S)}^\pm$), is used to store a $\pi$-polarized probe field ($\Omega_p^{\pi}$). At a later time, a second linearly polarized laser ($\Omega_{c(R)}^{\text{lin}}$), composed of two circularly polarized fields ($\Omega_{c(R)}^\pm$), is turned on to retrieve the stored excitation from the atomic medium. The control and the probe laser fields are all locked on resonance to the transition ($\ket{F=1}\longrightarrow\ket{F'=0}$.

\textit{Incoherent couplings:} The decay rates ($\gamma_{g}$ and $\gamma_{d}$) are engineered using two independent circularly polarized lasers fields that are tuned to transitions $\ket{F=1,m_F=\pm1}\longrightarrow\ket{F'=0,m_{F'}=0}$. These decay beams are kept at a constant red-detuning of 4 MHz red from this transition, even in presence of magnetic field, by driving their corresponding acousto-optic-modulators (AOMS) with independent frequency sources. This process ensures that the decay rates and their ratios remain constant as the magnetic field is varied along the quantization axis. Furthermore, the decay beams are counter-propagating to each other and are deliberately maintained at a small angle with respect to the control fields to ensure minimal coherence induced by these decay fields in the system. 

\textit{Classical phases:} The \textit{tone phase}, $\phi_c$, is the difference of the phase difference, $\phi_S-\phi_R$, between the storing ($\Omega_{c(S)}^\pm$) and retrieved $\Omega_{c(R)}^\pm$ control fields. The tone phase is varied by varying the linear polarization of the storing field with respect to the retrieving control fields.
The \textit{dynamic phase} ($\phi_d$) is varied by applying a magnetic field along the quantization axis with a pair of coils in Helmholtz configuration.

\textit{Optical pulses:}
All lasers are frequency locked either to atomic transitions or to frequency stabilized sources using beat-note techniques. VCOs are used for to tune the laser frequency dynamically while AOMs are used to create pulses or for turning fields on and off. In particular, the probe pulse width is set at $\sim$250 ns and the control field rise and fall ramp time at $\sim$ 100 ns. Amplitude, frequency and timing of lasers and magnetic fields are computer controlled, using a combination of  NI-DAQ system (PXIe-6738) and FPGA (XEM3001) for slow and fast timing control, respectively.

\begin{flushleft}
\textbf{S.II. Phase space representation (Husimi-Q function) of atomic spin-1 states:}
\end{flushleft}
\noindent
Of the four energy levels ($\ket{1}\equiv\ket{F=1,m_F=-1}$, $\ket{2}\equiv\ket{F=1,m_F=0}$, $\ket{3}\equiv\ket{F=1,m_F=1}$ and $\ket{4}\equiv\ket{F'=0,m_{F'}=0}$) that are directly involved in the dynamics, the excited state $\ket{4}$ is short lived with a life time of $\sim 30$ ns. The corresponding time scales involving probe pulse width and turn on and off times of control fields are significantly longer. Accordingly, most parts of the dynamics can be understood by adiabatically eliminating the excited state and considering the effective spin-1 three-level atom. We use the Husimi-Q function~\cite{HUSIMI40_s} to visualize dynamics of this effective spin-1 atom~\cite{Roulet18_s}. 
      \begin{figure*}[h!]
      \centering
       	\includegraphics[scale=0.3]{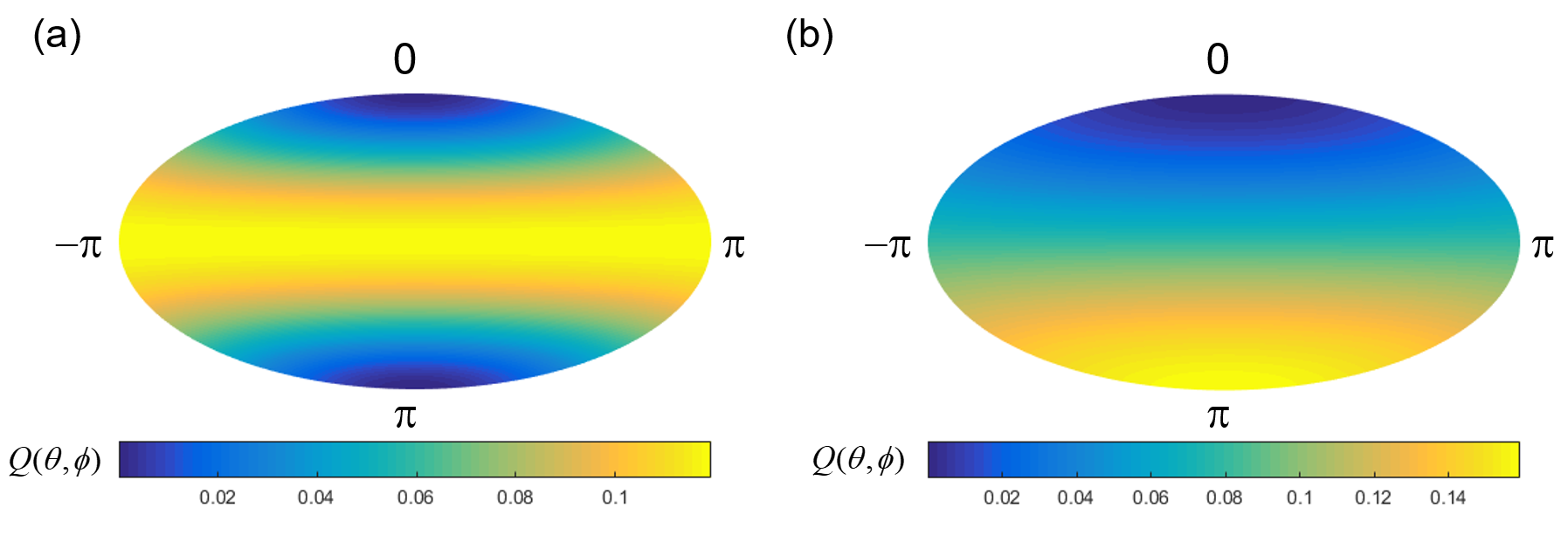}
       	\caption{(a) Equatorial limit cycle of spin-1 system with the population in state $\ket{F=1,m_F=0}$ or the state (0,1,0). (b) For comparison, a spin-1 state with population (0,$\frac{1}{3}$,$\frac{2}{3}$) are plotted. This limit cycle resembles a van der Pol oscillator deep in the quantum regime ~\cite{Sadeghpour13_s,Roulet19_s}.}
       	\label{fig:Limitcycle}
       \end{figure*} 
Husimi-Q function for a state $\hat{\rho}$ is defined by projecting it in a spin coherent state basis as:
 \begin{eqnarray}\label{HusimiQ}
 Q=\frac{3}{4\pi}\bra{\theta,\phi}\hat{\rho}\ket{\theta,\phi},
 \end{eqnarray}
 where $\ket{\theta,\phi}$ is the spin coherent state, and expressed as~\cite{Radcliffe71_s,Monica15_s}
 \begin{eqnarray}
 \ket{\theta,\phi}=\text{cos}^{2}\frac{\theta}{2}\left(\ket{1}+\sqrt{2}e^{i\phi}\text{tan}\left(\frac{\theta}{2}\right)\ket{2}+e^{i2\phi}\text{tan}^{2}\left(\frac{\theta}{2}\right)\ket{3}\right).
 \end{eqnarray}

\noindent 
Using the numerically simulated density matrix of the spin-1 three-level atom (section, S.IV.),  we plot this function in phase space parametrized by the two angles $\theta$ and $\phi$. 

To establish synchronization, one first needs to establish a limit cycle~\cite{Roulet18_s}. Such a limit cycle corresponds to a state with a closed trajectory in phase space. As proposed by Roulet \textit{et. al.}, in presence of anisotropic decays in spin-1 atoms, the limit cycle state gets localized and synchronizes itself to the tone phase at all times.
   
\bigskip
\noindent
\textbf{Limit cycle:}

\bigskip
\noindent

We initialize and prepare an atomic state, which when projected in phase space, forms an equatorial limit cycle corresponding to $\theta =\pi/2$, precessing about $z$ axis in  $\phi$. Such a state correspond to all atoms being prepared in state $\ket{2}$ and its Husimi-\textit{Q} function is $Q=\frac{3}{4\pi}\bra{\theta,\phi}\hat{\rho}\ket{\theta,\phi}=\frac{3}{8\pi}\text{sin}^{2}\theta$. In Fig.~\ref{fig:Limitcycle}(b), for comparison, we also plot the  Husimi-Q function for a van der Pol limit cycle oscillator deep in quantum regime~\cite{Walter2014QuantumSO_s} corresponding to the population $\hat{\rho}=\frac{1}{3}\ket{2}\bra{2}+\frac{2}{3}\ket{3}\bra{3}$.

 \begin{figure*}[h!]
  	\includegraphics[width=\linewidth]{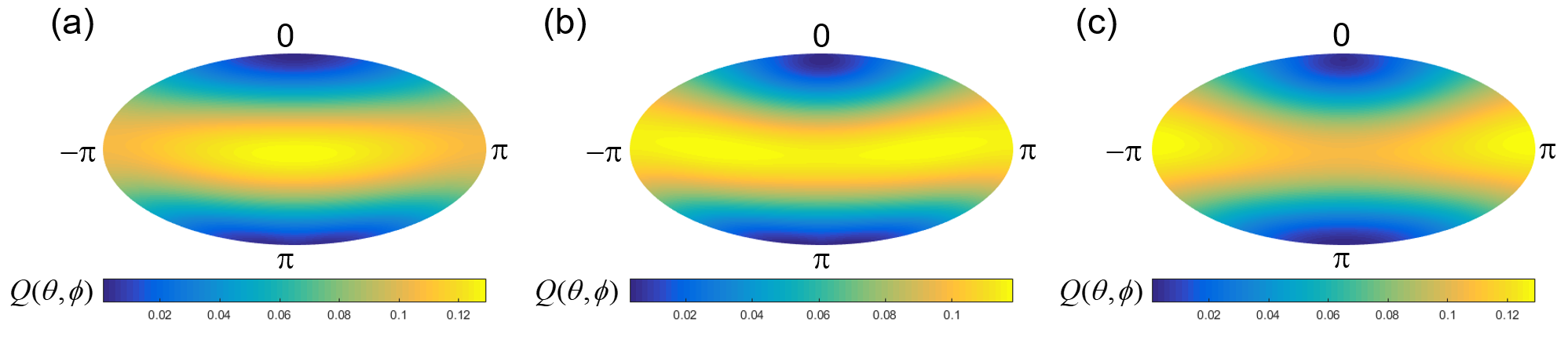}
  	\caption{(a) The phase distribution localized around $\phi=0$ in case of asymmetric decay $\gamma_d/\gamma_g=0.1$. (b) In the symmetric case $\gamma_d/\gamma_g=1$ there is no phase localization with slight distortion of the limit cycle. (c) Phase localized state for $\gamma_d/\gamma_g=10$, localized around $\phi=\pi$. 
  	}
  	\label{fig:PhaseLocking}
 \end{figure*}

\bigskip
\noindent
\textbf{Synchronization:}

\bigskip
\noindent

For synchronization we first stabilize the equatorial limit cycle state by introducing two decay channels $\gamma_g$ and $\gamma_d$, from states $\ket{1}$ and $\ket{3}$ to $\ket{2}$, respectively. Next we introduce the classical drive with a tone phase, which is the phase difference between the two tones $\eta_{-1,0}$ and $\eta_{0,1}$. The corresponding coherent couplings are generated using two photon transitions via an additional level $\ket{4}$, with control fields $\Omega_{c}^\pm$ and a probe $\Omega^\pi_p$ (Fig. 1, main text). When the control fields are adiabatically turned off, the probe pulse gets stored as DSPs. The storage of the probe pulse with two control fields effectively drives the spin-1 system and creates off-diagonal elements of the spin-1 density matrix, as DSPs.
 
Simulated Husimi-Q functions for evolving times, calculated from the density matrix elements obtained by solving optical Bloch equations, is plotted in Fig.~\ref{fig:PhaseLocking} using Hammer projection of a sphere~\cite{Goldberg07_s}. With storage of the probe, the limit cycle transforms to a phase localized state with two DSPs constructively interfering, with the tone phase $\phi_c=0$. This is particularly observable at finite detuning (magnetic field), such that, the phase localized state precess around z-axis with time. The precession leads to interference fringes in retrieved probe intensity at later times. However, when the two DSPs destructively interfere (for $\phi_c=\pi$), the initial limit cycle is only mildly perturbed with the \textit{Q} function uniformly distributed in the equatorial plane with phase synchronization (Fig.~\ref{fig:PhaseLocking}(b)). However, the scenario drastically alters when the two decay rates, $\gamma_g$ and $\gamma_d$ are made unequal. In such cases, the spin-1 atomic state, at the instance of storage, gets phase localized at $\phi=0$ and $\phi=\pi$ corresponding to asymmetric rates, with $\gamma_d/\gamma_g=0.1$ (Fig.~\ref{fig:PhaseLocking}(a)) or $10$ (Fig.~\ref{fig:PhaseLocking}(c)), respectively. Most importantly, even in presence of detuning the state remains locked to the tone and dynamic phase, over a range of dynamic phase. 

\bigskip
\noindent
\textbf{Synchronization function:}

\bigskip
\noindent 
	Although Husimi-Q function shows synchronization as a localization of the spin-1 state in phase space, it does not provide a quantitative measure. To quantify synchronization, Roulet et el.~\cite{Roulet18_s} introduced synchronization function $S(\phi)$, defined as 
    	\begin{eqnarray}\label{syncdef}
   			S(\phi)=\int_{0}^{\pi}Q(\theta,\phi)\sin\theta d\theta-\frac{1}{2\pi}.
    	\end{eqnarray}
    This function is non-zero for any synchronized state while it is identically zero for a limit-cycle state. 
    
    To estimate this function in context of our experiment, we first consider the equatorial limit-cycle state in presence of coherent couplings. The corresponding density matrix takes a form:   
	\begin{equation}
		\hat{\rho}=
		\begin{bmatrix}	
		0 & \rho_{12} & 0 \\
		\rho^{*}_{12} & 1 & \rho_{23} \\
		0 & \rho^{*}_{23} & 0\\
		\end{bmatrix},
		\label{eq:Perturbstate}
	\end{equation}
	where the off-diagonal terms are due to coherent, weak drives. 
	Husimi-Q function for this state can be estimated from equation~(\ref{HusimiQ}),as 
 	\begin{equation}\label{husimiQ}
 	 	Q({\theta,\phi})=\frac{3}{4\pi}\left<{\theta,\phi}|{\hat{\rho}}|{\theta,\phi}\right>=\frac{3}{4\pi}\left[{\alpha\sin\theta\cos^2\left({\frac{\theta}{2}}\right)+\beta\sin\theta\sin^2\left({\frac{\theta}{2}}\right)+\frac{1}{2}\sin^2\theta}\right],
 	\end{equation}
	 where
	  \begin{eqnarray}
	  \alpha=\frac{1}{\sqrt{2}}\left[{\rho^*_{12}\exp(-i\phi)+\rho_{12}\exp(i\phi)}\right],\nonumber\\
	  \beta=\frac{1}{\sqrt{2}}\left[{\rho^*_{23}\exp(-i\phi)+\rho_{23}\exp(i\phi)}\right].\nonumber
	  \end{eqnarray}
	In absence of coherent couplings ($\alpha=\beta=0$),i.e. for the original limit-cycle, the Husimi-Q representation is $Q({\theta,\phi})=\frac{3}{8\pi}\sin^2\theta$. When plotted in phase space (Fig.\ref{fig:Limitcycle}(a)) using Hammer projection of a sphere, $Q({\theta,\phi})$ is uniformly distributed with $\phi$ and centred around $\theta=\pi/2$. The synchronization function $S(\phi)$(equation~(\ref{syncdef})) in this case vanishes showing no sign of synchronization. 
	
	On the contrary, in presence of coherent couplings, off-diagonal terms of the density matrix (equation~(\ref{eq:Perturbstate})) are non-zero and the corresponding synchronization function is:
	 \begin{eqnarray}\label{syncfn}
    	 S(\phi)&=&\frac{3}{16}\left[\alpha+\beta\right].
 	\end{eqnarray}
		Substituting the values of $\alpha$ and $\beta$ in equation~(\ref{syncfn}) we get
	\begin{eqnarray}\label{syncfn1}
   		S(\phi)&=&\frac{3}{16\sqrt{2}}\left[{\rho^*_{12}\exp(-i\phi)+\rho_{12}\exp(i\phi)}+{\rho^*_{23}\exp(-i\phi)+\rho_{23}\exp(i\phi)}\right].
    \end{eqnarray}
Furthermore, due to an applied magnetic field ($B_z$) the ground state energy shifts by $\Delta_B=\mu_{B}B_z/2\hbar$. As a result each coherence acquire dynamic phase $\Delta_B\tau$ for a finite storage time $\tau$. Accordingly, $\rho_{12}$ and $\rho_{23}$ can be written as 
	\begin{eqnarray}\label{coherence}
        \rho_{12}&=&|\rho_{12}|e^{i{\Delta_B\tau}},\\
        \rho_{23}&=&|\rho_{23}|e^{i({\Delta_B\tau+\phi_c})},
     \end{eqnarray}
     where $\phi_c$ is the external tone phase.
Using equation (8) and (9) the synchronization function can be written as    
  	\begin{eqnarray}\label{syncfn2}
     		S(\phi)&=&\frac{3}{8\sqrt{2}}\left[{|\rho_{12}|\cos(\Delta_B\tau+\phi)+|\rho_{23}|\cos(\Delta_B\tau+\phi+\phi_c)}\right].
    \end{eqnarray}       
  Here, $\Delta_B$, $\tau$ and $\phi_c$ represent ground state energy shift due to an applied magnetic field along quantization axis, storage time and external tone phase, respectively.
  
  \noindent
 In equation~(\ref{syncfn2}), when the tones are out-of-phase ($\phi_c=\pi$), the synchronization function vanishes for all values of $\phi$ with no synchronization. However, in presence of asymmetric decay rates ($\gamma_g$,$\gamma_d$), one of the coherences dominate, along with the condition of destructive interference being lifted and the system synchronizing with at a finite $S(\phi)$. In presence of incoherent decays ($\gamma_g$ and $\gamma_d$), the coherences decay exponentially and take the form as	
     \begin{eqnarray}\label{coherenceDecay}
         \rho_{12}&=&|\rho_{12}|e^{-\gamma_g\tau}e^{i{\Delta_B\tau}},\\
         \rho_{23}&=&|\rho_{23}|e^{-\gamma_d\tau}e^{i({\Delta_B\tau+\phi_c})}.
      \end{eqnarray}
      Accordingly, the synchronization function in equation~(\ref{syncfn2}) becomes
        	\begin{eqnarray}\label{syncfn3}
           		S(\phi)&=&\frac{3}{8\sqrt{2}}\left[{|\rho_{12}|e^{-\gamma_g\tau}\cos(\Delta_B\tau+\phi)+|\rho_{23}|e^{-\gamma_d\tau}\cos(\Delta_B\tau+\phi+\phi_c)}\right].
          \end{eqnarray}

The maximum of synchronization function for the experimental and numeric data (main text, Fig. 2(f),2(g),2(h) and 2(i)) have been calculated by extracting the magnitude of generated coherences from the visibility of the interference fringes (main text, Fig. 2(a),2(b),2(c) and 2(d)). The retrieved pulse intensity ($I_R$) results from the interference of the two generated coherences, which is proportional to 
\begin{equation}
I_R \propto |\rho_{12}+\rho_{23}^*|^2.
\end{equation}
Accordingly, for particular $\Delta_B$, 
\begin{equation}\label{Rmax}
I_R^{max} \propto (|\rho_{12}|+|\rho_{23}|)^2,\\
\end{equation}
and
\begin{equation}\label{Rmin}
I_R^{min} \propto (|\rho_{12}|-|\rho_{23}|)^2.\\
\end{equation}
Using equations~(\ref{Rmax}) and ~(\ref{Rmin}) we can find $|\rho_{12}|$ and $|\rho_{23}|$ as,

\begin{eqnarray}
|\rho_{12}|=\kappa(\frac{\sqrt{I_R^{max}}+\sqrt{I_R^{min}}}{2}) \hspace{2mm} \text{and} \hspace{2mm} |\rho_{23}|=\kappa(\frac{\sqrt{I_R^{max}}-\sqrt{I_R^{min}}}{2}),
\end{eqnarray}
where $\kappa$ is a proportionality constant which depends on experimental parameter such as normalization, optical depth and retrieved powers of the system. Once we have $|\rho_{12}|$ and $|\rho_{23}|$, we can calculate the maximum of synchronization function using equation~(\ref{syncfn2}).

\begin{flushleft}
\textbf{S.III. Interference of dark state polaritons (DSPs): derivation of retrieved probe intensity}
\end{flushleft}

\noindent
Here we derive an analytic expression for the retrieved probe pulse intensity in a 4-level atomic system with three ground states. The derived expression for the retrieved probe intensity forms our primary basis for interpreting the experimental data of Figs. 1, 2 and 4 in main text. 

\bigskip
\noindent
\textbf{Interference of DSPs:}\label{DarkStatePolariton}

\bigskip
\noindent
Dark state polaritons are quasiparticles composed of a coherent superposition of atomic and photonic excitations and are usually deeply associated with the phenomenon of electro-magnetically induced transparency (EIT)~\cite{Fleischhauer00_s,Lukin01_s,Lukin02_s,Lukin03_s,Fleischhauer05_s}. These hybrid states propagates through the medium in a shape preserving manner and can be expressed as:
\begin{align}
	\hat{\psi}(z,t)&=\cos\theta(t)\hat{E}_p(z,t) - \sin\theta(t)\sqrt{N}\hat{\sigma}_{12}(z,t),\\
	\cos\theta(t)&=\frac{\Omega_c(t)}{\sqrt{|\Omega_c(t)|^2+g^2N}}, \hspace{5 mm} \sin\theta(t)=\frac{g\sqrt{N}}{\sqrt{|\Omega_c(t)|^2+g^2N}}.
	\label{eq:Polariton}
\end{align}
The state  $\hat{\psi}(z,t)$ in turn is governed by the equation of motion~\cite{Fleischhauer00_s}:
\begin{equation}
	\left[\frac{\partial}{\partial t}+c\cos^2\theta(t)\frac{\partial}{\partial z}\right]\hat{\psi}(z,t)=0,
\end{equation}
where $g$, $N$, $\hat{E}_p$ and $\sigma_{12}$ are the single atom cooperativity, total number of atoms, field operator and atomic operator, respectively. The generated DSP propagates with a group velocity $v_g(t)=c\cos^2\theta(t)$, rendering to a possibility of controlling its velocity by adiabatically changing the control Rabi frequency $\Omega_c$. In particular, when the control intensity is turned off, the probe pulse gets stored as superposition of atomic ground states or as atomic coherences. The atomic coherence evolves with time in the presence of a magnetic field, picking up a dynamic phase. One can retrieve the stored excitation back as a probe pulse, by adibatically turning on the control field at a later time. 

 We have extended this scheme to include two control fields and three ground states. A similar storage experiment results in two DSPs with relative phase difference between them. In particularly, we use two circularly polarized control fields to coupled the two edge states $\ket{1}\equiv\ket{F=1,m_F=-1}$ and $\ket{3}\equiv\ket{F=1,m_F=1}$ to the excited state $\ket{4}\equiv\ket{F'=0,m_{F'}=0}$, and a $\pi$-polarized probe field that coupled the central groubd state $\ket{2}\equiv\ket{F=1,m_F=0}$ to the excited state $\ket{4}$. The fields have Rabi frequencies $\left(\Omega_{c(S)}^{+},\Omega_{c(S)}^{-}\right)$, and $\left(\Omega_{p}^{\pi}\right)$, respectively. These fields result in two DSPs with corresponding wave functions:~\cite{Xiao11_s,Martin08_s}  
\begin{equation}
\psi^+(z,t)=\text{cos}\theta_1(t)\Omega_{p}^{\pi}(z,t)-\text{sin}\theta_1(t)\sqrt{N}\rho_{12}(z,t)\\
\label{eq:Polariton1}
\end{equation}
\begin{equation}
\psi^-(z,t)=\text{cos}\theta_2(t)\Omega_{p}^{\pi}(z,t)-\text{sin}\theta_2(t)\sqrt{N}\rho_{23}^*(z,t),
\label{eq:Polariton2}
\end{equation}  
with corresponding atomic coherences $\rho_{12}$ and $\rho_{23}$. 

 \begin{figure*}
 \centering
 	\includegraphics[scale=0.7]{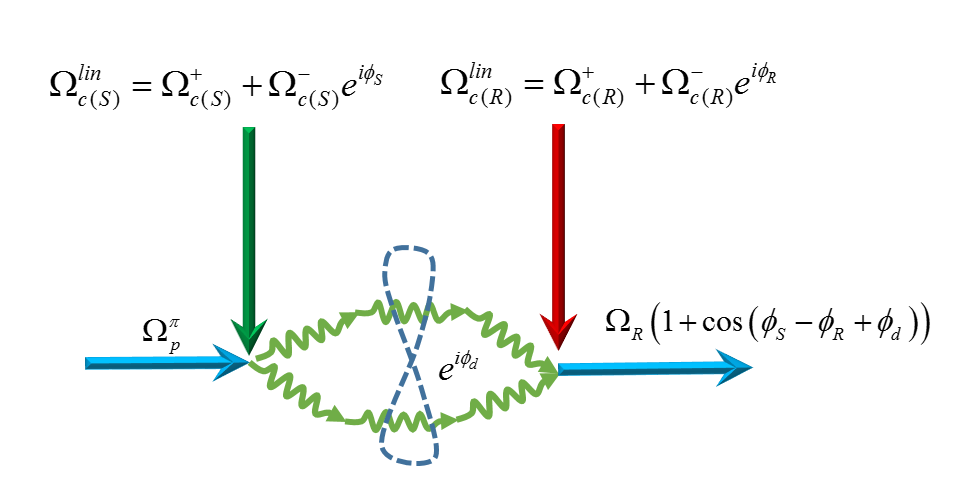}
 	\caption{A diagrammatic representation of the equivalent atom-photon interferometer, corresponding to the experiments reported here, with two stored DSPs (shown with green curvy arrows). All other symbols have usual meaning as defined in the main text. The output of the interferometer depends on the relative dynamic phase difference ($\phi_{S}$-$\phi_{R}$+$\phi_{d}$) of the two polaritons.}
 	\label{fig:Interferrometer}
 \end{figure*}
 
\noindent 
The mixing angles are expressed as:
\begin{eqnarray}
\cos\theta_1(t)=\frac{\Omega_{c(S)}^{+}(t)}{\sqrt{|\Omega_{c(S)}^{+}(t)|^{2}+g^2N}} \text{, and }\sin\theta_1(t)=\frac{g\sqrt{N}}{\sqrt{|\Omega_{c(S)}^{+}(t)|^{2}+g^2N}}\\
\cos\theta_2(t)=\frac{\Omega_{c(S)}^{-}(t)}{\sqrt{|\Omega_{c(S)}^{-}(t)|^{2}+g^2N}} \text{, and }\sin\theta_2(t)=\frac{g\sqrt{N}}{\sqrt{|\Omega_{c(S)}^{-}(t)|^{2}+g^2N}}
\end{eqnarray}
By adiabatically turning off the control fields $\Omega_{c(S)}^\pm$, the dark state polaritons given in equations~(\ref{eq:Polariton1}), and (\ref{eq:Polariton2}) get stored in the form of atomic coherences ($\rho_{12}, \text{and}~\rho_{23}$). While stored, the coherences evolve with time and pick up equal and opposite dynamic phases in presence of a magnetic field along the quantization axis. At a later time when the control fields $\left(\Omega_{c(R)}^{+},\Omega_{c(R)}^{-}\right)$ are turned back on, the atomic coherences are mapped back onto an optical pulse in the original probe pulse direction. 

The corresponding retrieved optical pulse ($\Omega_R$) can thereby be expressed as:
\begin{align}
\Omega_R(t+\tau)&\propto\sqrt{N}\left[\text{cos}\theta_3(t)\rho_{12}(z,t+\tau)+\text{cos}\theta_4(t)\rho_{23}^*(z,t+\tau)\right]\\
&=\sqrt{N}e^{-\tau^2\Gamma_d^2/4}\left[\text{cos}\theta_3(t)\rho_{12}(z,t)e^{-i\phi_-}+\text{cos}\theta_4(t)\rho_{23}^*(z,t)e^{-i\phi_+}\right],
\label{eq:RetreivedPulse}
\end{align} 
where $\Gamma_d$ is the Doppler timescale corresponding to a Gaussian decay while the time evolution of $\ket{1}$ and $\ket{3}$ in presence of magnetic field $B_z$ (along the quantization axis) results in the dynamic phase 
$\phi_\pm={\frac{\mu_{B}m_{F}g_{F}B_z\tau}{\hbar}}$. Here $\mu_{B}$,$m_{F}$, and $g_{F}$ are Bohr magneton, magnetic quantum number of the respective state, and hyperfine Land\'{e} g-factor, respectively. The total relative dynamic phase is then $\phi_d=\phi_--\phi_+=\frac{\mu_{B}B_z\tau}{{\hbar}}=2\Delta_B\tau$, where $g_F=-\frac{1}{2}$ and $\Delta_B=\frac{\mu_{B}B_z}{{2\hbar}}$.

In equation~(\ref{eq:RetreivedPulse}) the weights of the contributions from the two coherences in the retrieved pulse are 
\begin{eqnarray}
\text{cos}\theta_3(t)=\frac{\Omega_{c(R)}^{+}(t)}{\sqrt{|\Omega_{c(R)}^{+}(t)|^{2}+g^2N}} \text{, and }\text{cos}\theta_4(t)=\frac{\Omega_{c(R)}^{-}(t)}{\sqrt{|\Omega_{c(R)}^{-}(t)|^{2}+g^2N}}.
\end{eqnarray} 
Also, $|\rho_{12}(z,t)|\propto|\Omega_{c(S)}^{+}(z,t)|$ and $|\rho_{23}(z,t)|\propto|\Omega_{c(S)}^{-}(z,t)|$. For the special case, $|\Omega_{c(R)}^{+}(t)|=|\Omega_{c(R)}^{-}(t)|=|\Omega_{c(S)}^{+}(t)|=|\Omega_{c(S)}^{-}(t)|=|\Omega_{c}(t)|$,
the recovery pulse intensity can be expressed as, 
\begin{eqnarray}
\boxed{I_R\propto|\Omega_R(t+\tau)|^{2}\propto |\Omega_{c}|^{2}e^{-\tau^2\Gamma_d^2/2}\cos^{2}\frac{1}{2}(\phi_c+2\Delta_B\tau)}.
\end{eqnarray}
\noindent
Here $\phi_c=\phi_S-\phi_R$, where $\phi_{S}$ is the phase difference between $\Omega_{c(S)}^+$ and $\Omega_{c(S)}^-$, and $\phi_{R}$ between $\Omega_{c(R)}^+$ and $\Omega_{c(R)}^-$ (Fig.~\ref{fig:Interferrometer}). 
In our experiments we kept $\phi_R=0$, so that $\phi_c=\phi_S$ becomes the tone phase. 
Fig.~\ref{fig:Interferrometer} depicts a diagrammatic way of representing interference of the two DSPs in the medium. The experiment effectively can be represented as an atom-photon interferometer, with the stored DSPs (light green curvy arrow) are generated in the medium when a linearly polarized control field $\Omega_{c(S)}^\text{lin}$ (green arrow, composed of two circularly polarized light $\Omega_{c(S)}^\pm$) is adiabatically turned off. Generation of the DSPs thereby constitute the first ``beam splitter" while the two stored DSPs constitutes the two arms of the interferometer. In presence of a magnetic field ($B_z$, along the quantization axis), the two polaritons acquire equal and opposite dynamic phases. The resulting relative phase difference of the two arms of the interferometer leaves its signture in the retrieved pulse intensity, using a second linearly polarized control field $\Omega_{c(R)}^\text{lin}$ (red arrow, composed of two circularly polarized light $\Omega_{c(R)}^\pm$). This second control field constitute the second ``beam splitter".
While the DSPs are stored in atoms, in presence of anisotropic decay channels $\gamma_g$ and $\gamma_d$, the relative phase of the interferometer gets \textit{locked} to the total phase that includes the relative phase of the beam splitters (the phase difference between the two sets of storing and retrieving control fields) and the phase due to an applied magnetic field. This is the primary physical manifestation of quantum synchronization.

 \begin{figure*}
 \centering
 	\includegraphics[scale=0.45]{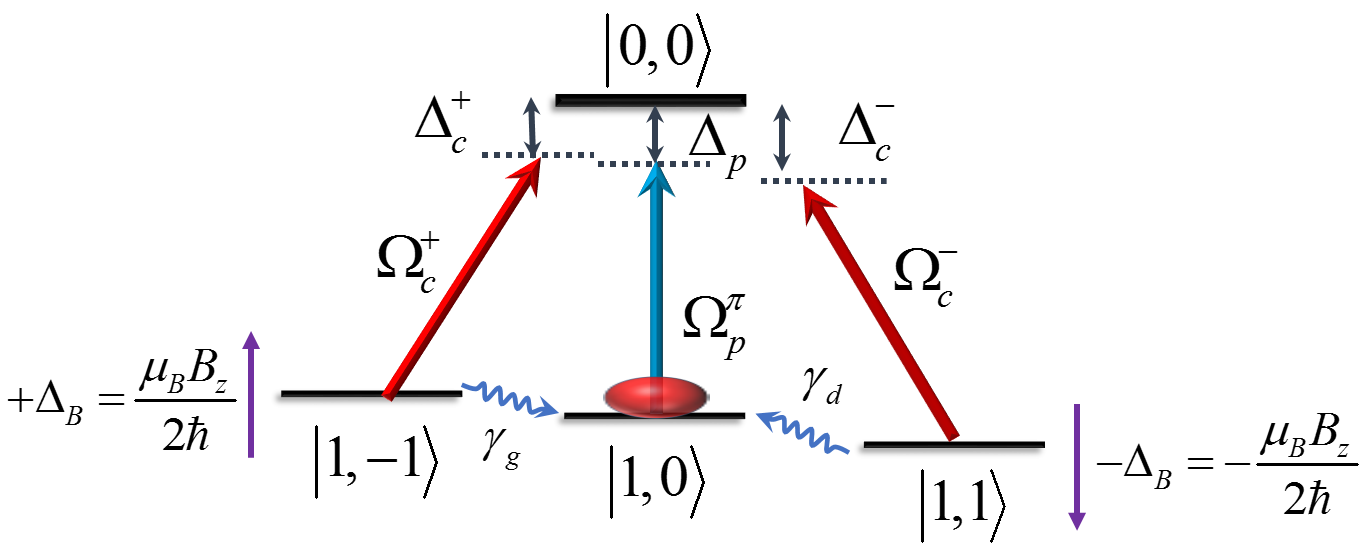}
 	\caption{Energy level diagram of a four level atom. Here control ($\Omega_c^\pm$, red) and probe fields ($\Omega_{p}^\pi$, blue) connect the ground state $\ket{1,\pm1}$ and $\ket{1,0}$ to the excited state $\ket{0,0}$, respectively. The corresponding detuning for control and probe field are $\Delta_{c}^\pm$ and $\Delta_p$, respectively. The incoherent decays $\gamma_g$ and $\gamma_d$ are introduced from the ground states $\ket{1,-1}$ and $\ket{1,1}$ to $\ket{1,0}$, respectively. $\Delta_B$ is the ground state energy shift due to presence of a magnetic field $B_z$ and given by $\Delta_B=\mu_BB_z/2\hbar$, where $\mu_B$ and $\hbar$ correspond to Bohr magneton and reduced Plank constant.} 
 	\label{fig:leveldiag}
 \end{figure*}
 
\begin{flushleft}
\textbf{S.IV. Numerical simulations:}
\end{flushleft}
\noindent

We numerically solve optical Bloch equations and simulate time dynamics of the probe pulse in presence and absence of a control field. We observe numerically simulated probe transmission traces to be in excellent agreement with experiments (Fig.~\ref{fig:ExpSimulation}). We use this correspondence to reconstruct the density matrix of experimental spin-1 atoms and plot the corresponding state projection (Husimi-Q function) in phase space.

\bigskip
\noindent
{\textbf{Optical Bloch equations and density matrix elements:}}

\bigskip
\noindent

To simulate our experiment, we construct a 4-level atomic system with three ground states $\ket{1},\ket{2},\ket{3}$ and an excited state $\ket{4}$. These ground and excited states correspond to $\ket{F=1,m_F=-1},\ket{F=1,m_F=0},\ket{F=1,m_F=1}$ hyperfine ground states and $\ket{F'=0,m_{F'}=0}$ excited states (Fig.~\ref{fig:leveldiag}), corresponding to D2 transition of $^{87}Rb$ atoms. We denote them as, $\ket{1}\equiv\ket{F=1,m_F=-1}$, $\ket{2}\equiv\ket{F=1,m_F=0}$, $\ket{3}\equiv\ket{F=1,m_F=1}$, and $\ket{4}\equiv\ket{F'=0,m_{F'}=0}$ and we consider the atoms to be independent entities. 

The effective Hamiltonian of such a four-level system can be expressed, in a rotating frame, as: 
\begin{equation}
\hat{H}=-\hbar
\begin{bmatrix}	
(\Delta_p-\Delta_{c}^{+}+\Delta_B) & 0 & 0 & \Omega_{c}'^{+}\\
0 & 0 & 0 & \Omega_p'^\pi\\
0 & 0 & (\Delta_p-\Delta_{c}^{-}-\Delta_B) & \Omega_{c}'^{-}\\
\left(\Omega_{c}'^{+}\right)^{*} & (\Omega_p'^\pi)^* & \left(\Omega_{c}'^{-}\right)^{*} & \Delta_p
\end{bmatrix}.
\label{eq:4LevelHamiltonian}
\end{equation}

\noindent
Here the control Rabi frequency at different times are defined as:
\begin{equation}
\Omega_{c}'^{\pm}(z,t) =
\left\{
	\begin{array}{ll}
		\Omega_{c(S)}'^{\pm}(z,0)e^{-(t-t_{on1})^2/2\tau_c^2}e^{i\phi_{S}^{\pm}}  & \mbox{if } t\leq t_{on1} \\
		\Omega_{c(S)}'^{\pm}(z,0)e^{i\phi_{S}^{\pm}} & \mbox{if } t_{on1}<t\leq t_{off}\\
			\Omega_{c(S)}'^{\pm}(z,0)e^{-(t-t_{off})^2/2\tau_c^2}e^{i\phi_{S}^{\pm}}+\Omega_{c(R)}'^{\pm}(z,0)e^{-(t-t_{on2})^2/2\tau_c^2} & \mbox{if } t_{off}<t\leq t_{on2}\\
		\Omega_{c(R)}'^{\pm}(z,0) & \mbox{if } t>t_{on2}
	\end{array},
\right.
\end{equation}

 \noindent
 where $\tau_c=70$ ns is the ramp time corresponding to turn on and off of the storing (retrieving) control fields, $\Omega_{c(S)}'^{+}$ ($\Omega_{c(R)}'^{+}$) and $\Omega_{c(S)}'^{-}$ ($\Omega_{c(R)}'^{-}$), connect the levels $\ket{1}$ and $\ket{3}$ to $\ket{4}$, respectively (Fig. 1, main text). The difference of the two control field phases, $\phi_c=\phi_{S}^{+}-\phi_{S}^{-}$, is what we refer to in main text as the tone phase. The phase difference of the coherences of the spin-1 atoms i.e. the phase difference between the two DSPs locks to this classical tone phase when synchronized in presence of anisotropic decay. 
 
 At time instance $t_{on1}\sim t_I+2\tau_c$, when the control fields ($\Omega_{c(S)}'^{\pm}$) are (adiabatically) turned on and in presence of a probe pulse, two DSPs form in the system. A part of the DSPs is electromagnetic field dependent while the other part is atomic coherence. When the control fields are (adiabatically) turned off at time instance $t_{off} \sim t_{II}$, the DSPs are purely atomic and get stored in the form of ground state coherences ($\rho_{12}$ and $\rho_{23}$). This stored excitations are converted back into electromagnetic fields, when the control fields with Rabi frequencies $\Omega_{c(R)}'^{\pm}(z,0)=f\Omega_{c(S)}'^{\pm}(z,0)$,  amplified by a factor \textit{f}, are turned back on at time instance $t_{on2} \sim t_{IV}$. The time instances $t_{I}, t_{II}~\text{and}~t_{IV}$ are shown in Fig. 1(c), main text. The corresponding two atomic coherences, when mapped back photonic state, interfere and the phase of interference varies with either varying dynamics phase (by applied magnetic field) or with varying tone phase i.e. the difference of the phase the two circularly polarized control fields. Experimentally, we change the tone phase using a half wave plate in the path of the control fields. 
 
 Numerically, we set the turn on and turn off times for control fields at $t_{on1}=0.4$ $\mu$s, $t_{off}=1.91$ $\mu$s, while the storage time is varied by varying $t_{on2}$ for the following experiments (Fig.~\ref{fig:ExpSimulation}). The probe Rabi frequency is defined as 
 \begin{equation}
 \Omega_{p}'^\pi(z,t) =
	\begin{array}{ll}
		\Omega_{p}'^\pi(z,0)e^{-(t-t_p)^2/2\tau_p^2}  & \mbox{} \forall t \\
	\end{array}
 \end{equation}

 \noindent
 We turn on the probe pulse at $t_p=1.90$ $\mu$s. It has a pulse width $\tau_p=$ 200 ns. The probe field connects $\ket{2}\longrightarrow\ket{4}$ transition. The Rabi frequencies of the control fields are defined as $\Omega_{c,p}^{\pm,\pi}=d^{\pm,\pi}_{c,p}\cdot\mathcal{E}^{\pm,\pi}_{c,p}/\hbar=2\Omega'^{\pm,\pi}_{c,p}$, where $d^{\pm,\pi}_{c,p}$ correspond to the transition dipole moment, and $\mathcal{E}^{\pm,\pi}_{c,p}$ electric field amplitude for the corresponding fields. The detuning of the fields from atomic transition energies are $\Delta^{+}_{c}=\omega^{+}_{c}-\omega_{14}$, $\Delta^{-}_{c}=\omega^{-}_{c}-\omega_{34}$, and $\Delta_{p}=\omega_p-\omega_{24}$. Here $\omega^{+}_{c},\omega^{-}_{c}$, and $\omega_p$ correspond to respective laser frequencies. To incorporate effect of the applied magnetic field (along the quantization axis), we add energy shifts $\Delta_B$ to ground states $\ket{1}$ and $\ket{3}$ (Fig.~\ref{fig:leveldiag}) to simulate the effect of external magnetic field ($B_z$). 

\begin{figure*}
	\includegraphics[width=\linewidth]{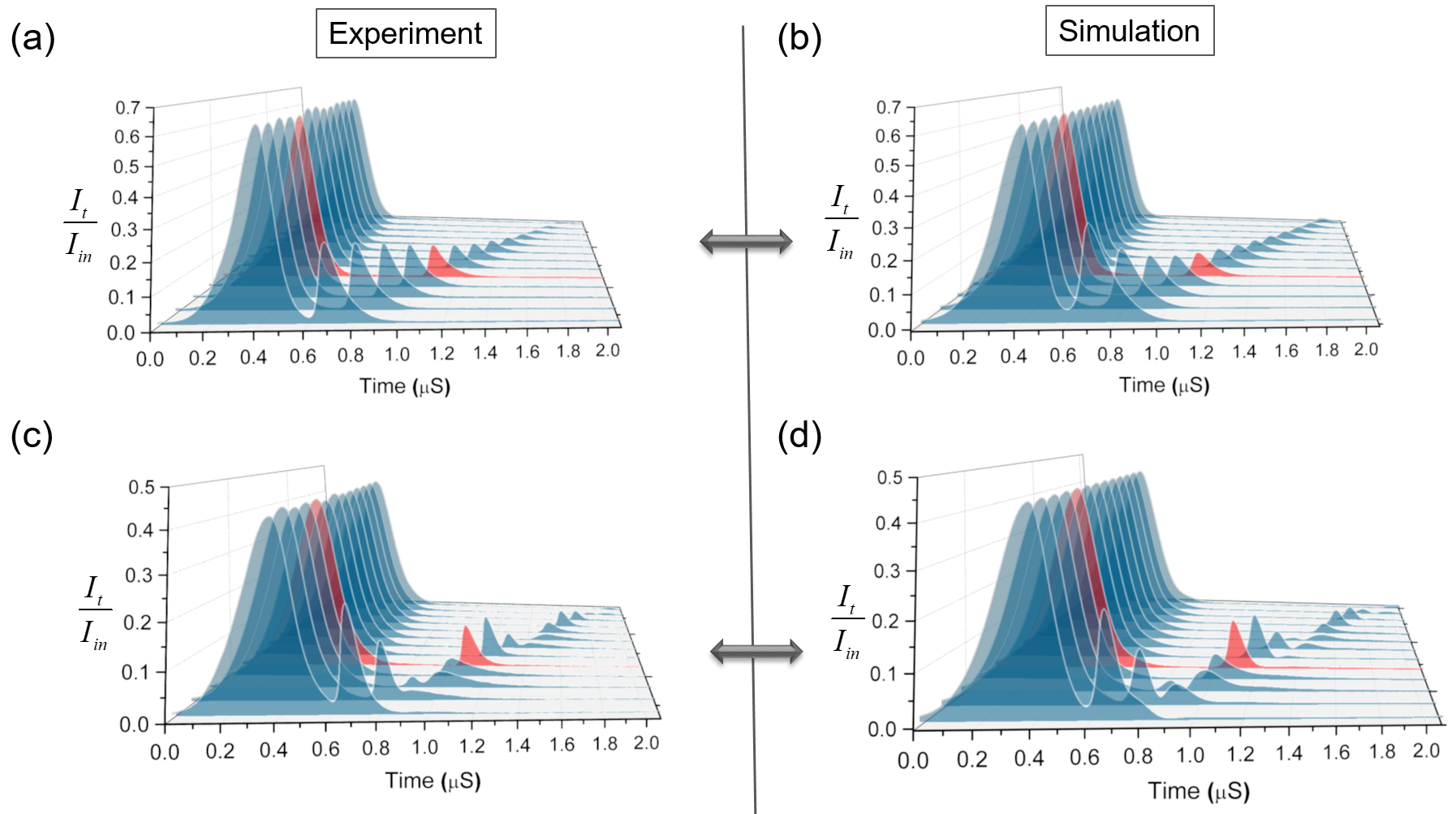}
	\caption{(a) and (c) Experimental plots of stored and retrieved probe pulse with varying storage time $\tau$ for zero magnetic field ($\Delta_B=0$) and $\Delta_B=1.34$ MHz, respectively. (b) and (d) simulated plots of stored and retrieval probe pulse to simulate the probe pulse obtained in experiments corresponding to (a) and (c) with $\Delta_B=0$ and 1.05 MHz, respectively.  Here I$_p$ and I$_0$ are transmitted and input probe field intensities. Parameters for simulations are as tabulated in Table S1.}
	\label{fig:ExpSimulation} 
\end{figure*}
 
%
 
Dynamics of the atomic density matrix of the the 4-level atom is governed by a master equation which can be expressed as: 
\begin{align}
\dot{\hat{\rho}}(t)=-\frac{i}{\hbar}[\hat{H},\hat{\rho}]+\hat{\mathcal{L}}(\rho).
\label{eq:MasterEqn}
\end{align}
where $H$ stands for the system Hamiltonian (eqn.~(\ref{eq:4LevelHamiltonian})), and $\hat{\mathcal{L}}(\rho)$ includes all the decays. We numerically integrate the master equation (\ref{eq:MasterEqn}). The corresponding dynamical equations for elements of the density matrix  are:
 \begin{align}
 \dot{\rho}_{11}=&-i[\rho_{14}^{*}(\Omega_{c}'^{+}){*}-\rho_{14}\Omega_{c}'^{+}]+\Gamma_{41}\rho_{44}-\gamma_{g}\rho_{11}-\gamma_{th}(\rho_{11}-\rho_{11}^{eq})\\
 \dot{\rho}_{12}=&-i[\rho_{12}(\Delta_p-\Delta_{c}^{+}+\Delta_B)+\rho_{24}^{*}(\Omega_{c}'^{+})^{*}-\rho_{14}\Omega_{p}'^\pi]-(\gamma_{12}+\gamma_{th})\rho_{12}\\
 \dot{\rho}_{13}=&-i[\rho_{13}(\Delta_{c}^{-}-\Delta_{c}^{+}+2\Delta_B)+\rho_{34}^{*}(\Omega_{c}'^{+})^{*}-\rho_{14}\Omega_{c}'^{-}]-(\gamma_{13}+\gamma_{th})\rho_{13}\\
 \dot{\rho}_{14}=&-i[(\Omega_{c}'^{+})^{*}(\rho_{44}-\rho_{11})-\rho_{12}(\Omega_{p}'^\pi)^{*}+\Delta_B\rho_{14}-\rho_{13}(\Omega_{c}'^{-})^{*}-\Delta_{c}^{+}\rho_{14}]-(\gamma_{41}+\gamma_{th})\rho_{14}\\
 \dot{\rho}_{22}=&-i[\rho_{24}^{*}(\Omega_{p}'^\pi)^{*}-\rho_{24}\Omega_{p}'^\pi]+\Gamma_{42}\rho_{44}+\gamma_{g}\rho_{11}+\gamma_{d}\rho_{33}-\gamma_{th}(\rho_{22}-\rho_{22}^{eq})\\
 \dot{\rho}_{23}=&-i[\rho_{23}(\Delta_{c}^{-}-\Delta_{p}+\Delta_B)+\rho_{34}^{*}(\Omega_{p}'^\pi)^{*}-\rho_{24}\Omega_{c}'^{-}]-(\gamma_{23}+\gamma_{th})\rho_{23}\\
 \dot{\rho}_{24}=&-i[(\Omega_{p}'^\pi)^{*}(\rho_{44}-\rho_{22})-\rho_{23}(\Omega_{c}'^{-})^{*}-\rho_{12}^{*}(\Omega_{c}'^{+})^{*}-\Delta_{p}\rho_{24}]-(\gamma_{42}+\gamma_{th})\rho_{24}\\
 \dot{\rho}_{33}=&-i[\rho_{34}^{*}(\Omega_{c}'^{-})^{*}-\rho_{34}\Omega_{c}'^{-}]+\Gamma_{43}\rho_{44}-\gamma_{d}\rho_{33}-\gamma_{th}(\rho_{33}-\rho_{33}^{eq})\\
 \dot{\rho}_{34}=&-i[(\Omega_{c}'^{-})^{*}(\rho_{44}-\rho_{33})-\rho_{13}^{*}(\Omega_{c}'^{+})^{*}-\rho_{23}^{*}(\Omega_{p}'^\pi)^{*}-\Delta_B\rho_{34}-\Delta_{c}^{-}\rho_{34}]-(\gamma_{43}+\gamma_{th})\rho_{34}\\
 \dot{\rho}_{44}=&-i[-\rho_{34}^{*}(\Omega_{c}'^{-})^{*}+\rho_{34}\Omega_{c}'^{-}-\rho_{14}^{*}(\Omega_{c}'^{+})^{*}+\rho_{14}\Omega_{c}'^{+}-\rho_{24}^{*}(\Omega_{p}'^\pi)^{*}+\rho_{24}\Omega_{p}'^\pi]\nonumber\\
 &-\Gamma_{44}\rho_{44}-\gamma_{th}(\rho_{44}-\rho_{44}^{eq}).
 \label{eq:OpticBloch}
 \end{align}
\noindent
Here $\Gamma_{41},\Gamma_{42}, ~\text{and}~\Gamma_{43}$ are the effective radiative decays from $\ket{4}$ to $\ket{1}$, $\ket{2}$, and $\ket{3}$, respectively, and are expressed as: $\Gamma_{41}=\Gamma_{42}=\Gamma_{43}=\Gamma_{44}/3$. The rate $\gamma_{th}$ is used for Doppler decoherence and is taken as thermal transit time of the atoms through the transverse cross-section of the probe beam. $\gamma_{g}$ and $\gamma_{d}$ correspond to the engineered incoherent decay from $\ket{1}$ and $\ket{3}$ to $\ket{2}$, respectively.
\begin{align}
\gamma_{12}&=\gamma_{g}/2+\gamma_{\text{dec}}\\
\gamma_{13}&=\gamma_{g}/2+\gamma_{d}/2+\gamma_{\text{dec}}\\
\gamma_{23}&=\gamma_{d}/2+\gamma_{\text{dec}}\\
\gamma_{43}&=\gamma_{d}/2+\Gamma_{44}/2+\gamma_{\text{dec}}\\
\gamma_{42}&=\Gamma_{44}/2+\gamma_{\text{dec}}\\
\gamma_{41}&=\Gamma_{44}/2+\gamma_{g}/2+\gamma_{\text{dec}},
\end{align}
where $\gamma_{\text{dec}}$ is a decoherence due to other effects including stray, oscillating magnetic fields, vacuum impurity, and wave-vector phase mismatch due to angles between control and probe fields. 

We simultaneously and self-consistently integrate the corresponding Maxwell equation to obtain dynamical evolution of the propagating probe field. Using slowly-varying-envelope-approximation (SVEA), the corresponding wave equations for fields can be expressed as:
\begin{align}
	\frac{1}{c}\frac{\partial}{\partial t}\Omega_{c}'^{+}(z,t)&+\frac{\partial}{\partial z}\Omega_{c}'^{+}(z,t)=-i\mu_{a}\rho_{14}(z,t)\\
	\frac{1}{c}\frac{\partial}{\partial t}\Omega_{c}'^{-}(z,t)&+\frac{\partial}{\partial z}\Omega_{c}'^{-}(z,t)=-i\mu_{a}\rho_{34}(z,t)\\
	\frac{1}{c}\frac{\partial}{\partial t}\Omega_{p}'^\pi(z,t)&+\frac{\partial}{\partial z}\Omega_{p}'^\pi(z,t)=-i\mu_{a}\rho_{24}(z,t).
\end{align}
Here $\mu_{a}=\frac{Nd^2\omega}{\hbar\epsilon_0}$ where N, d, $\omega$, and $\epsilon_0$ correspond to number of atoms, dipole moment, field frequencies and free space permittivity, respectively. The above equation can further be simplified by introducing the new variables $z=\zeta$ and $t'=t-z/c$~\cite{Eberly94_s}. The wave operator $\partial/\partial z+\partial/\partial ct$ simply takes the form under this transformation as $\partial/\partial\zeta$, and the propagation equation of the fields becomes
\begin{align}
	\frac{\partial}{\partial \zeta}\Omega_{c}'^{+}(z,t)(\zeta)&=-i\mu_{a}\rho_{14}(\zeta)\\
	\frac{\partial}{\partial \zeta}\Omega_{c}'^{-}(z,t)(\zeta)&=-i\mu_{a}\rho_{34}(\zeta)\\
	\frac{\partial}{\partial \zeta}\Omega_{p}'^\pi(z,t)(\zeta)&=-i\mu_{a}\rho_{24}(\zeta).	
\end{align}

From the propagation of equation of probe field we obtain the transmitted probe field as $\Omega_p'^\pi(\zeta)=\Omega_p'^\pi(\zeta-d\zeta)+\int_{0}^{L}\alpha\times Im(\rho_{24}(\zeta))d\zeta$, where $L$ is the length of the medium and $\alpha$ is a constant~\cite{arif16_s,arif18_s}. Fig.~\ref{fig:ExpSimulation}(a) and \ref{fig:ExpSimulation}(c) show the experimental traces obtained for stored and retrieved probe pulses for varying storage times $\tau$ at $\Delta_B=0$ and $1.34$ MHz, respectively. As we increase the storage time, the retrieved pulse intensity decreases monotonically due to Doppler and other decoherences present in the system. For non zero detuning, an additional dynamic phase $2\Delta_B\tau$ develops (S.III. Interference of dark state polaritons), and the retrieved pulse intensity develop fringes in time due to the acquired dynamic phase (Fig.~\ref{fig:ExpSimulation}(c)). The simulated traces are fitted to experimental traces (Fig.~\ref{fig:ExpSimulation}) by judiciously choosing and tuning the free parameters in the optical Bloch equations~(5)-(14). The simulated traces are not Doppler averaged over the velocity distribution of the atoms. Accordingly, we observe exponential decay of retrieved pulse intensity with increasing storage time. On the contrary, in experiments, we observe Gaussian decay of retrieved pulse and we ascribe the resulting Gaussianity to the essential thermal nature of the ensemble of cold, trapped atoms. The value of the parameters in simulation are listed in Table~S1.
 
 \begin{table}[t]
 	\begin{tabular}{|c||c|c|c|c|c|c|c|c|c|} 
 		\hline
 		Parameters & $\Gamma_{44}$ & $\gamma_{th}$ &  $\gamma_{dec}$ & $|\Omega_{c(S)}^{+}|$ & $|\Omega_{c(S)}^{-}|$ & $|\Omega_{p}^\pi|$ & $\gamma_{g}$ & $\Delta_B$ & \textit{f}\\
 		\hline 
 		Values & $(2\pi)~6$ MHz & $(2\pi)~0.01$ MHz & $(2\pi)~0.15$ MHz & $0.95~\Gamma_{44}$ & $0.95~\Gamma_{44}$ & $0.55~\Gamma_{44}$ & $(2\pi)~0.20$ MHz & $(2\pi)~1.05$ MHz & 1.25\\  
 		\hline
 	\end{tabular} 
 	\caption{Parameters used in numerical simulations.}
 	\label{Tab:parameter}
 \end{table}

\begin{flushleft}
\textbf{S.V. Experimental Details:}
\end{flushleft}
\noindent
In this section we discuss details of some of the experimental techniques used. These include a discussion of experimental timing sequence of events, estimation of temperature of the laser-cooled atoms, calibration of magnetic field, and methods of engineering incoherent decay rates.

 \begin{figure*}
 	\includegraphics[width=\linewidth]{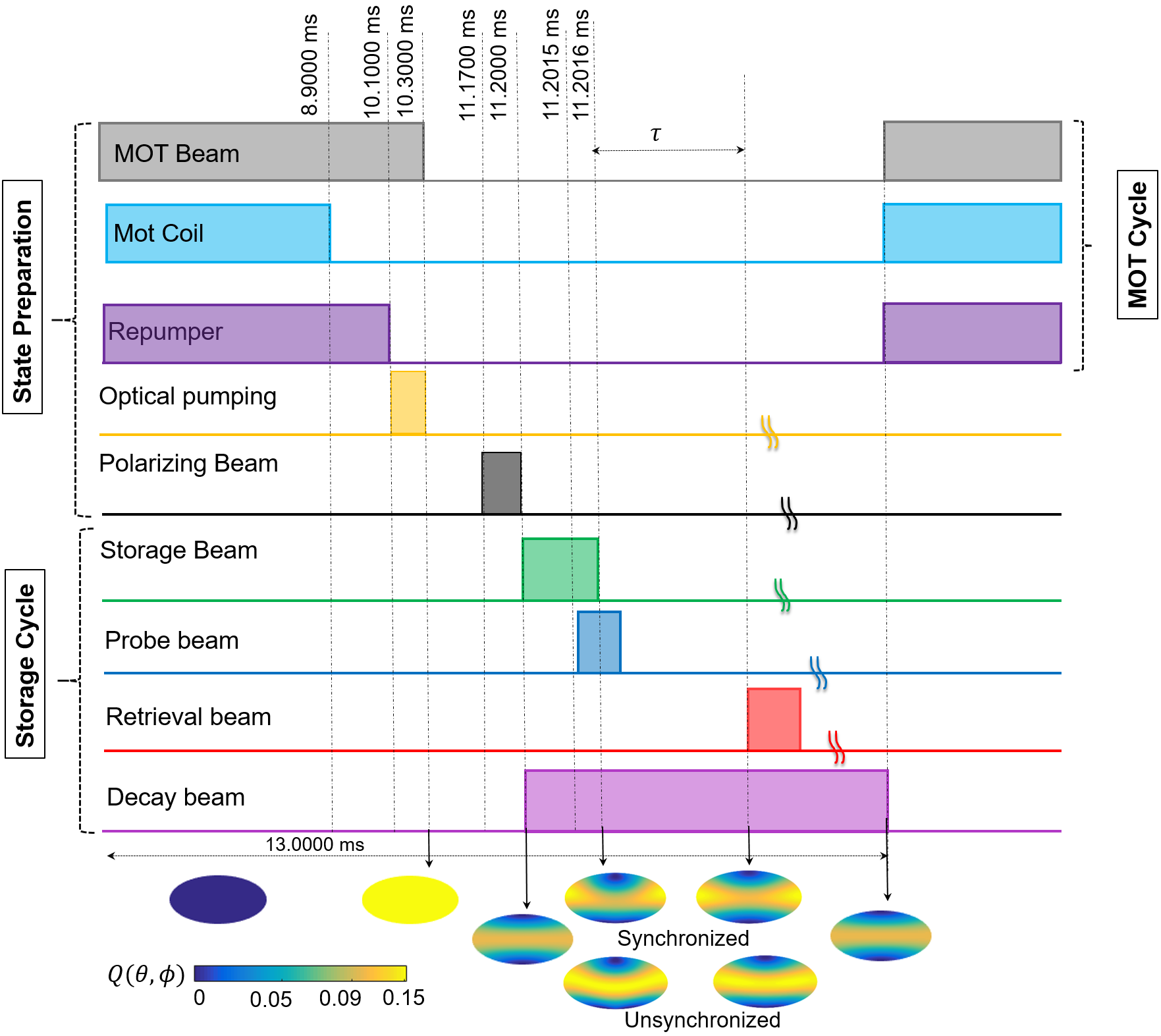}
 	\caption{A cartoon of timing sequence for the entire experiment, including state preparation, storage cycle comprising of control, probe and decay fields. For reference, the corresponding numerically simulated spin-1 atomic state, represented by Husimi-Q functions, are shown at different time instances of the experiment.
 	}
 	\label{fig:Motcycle} 
 \end{figure*}

\bigskip
\noindent
\textbf{(a) MOT Cycle:}

\bigskip
\noindent
Fig.~\ref{fig:Motcycle} shows a cartoon diagram corresponding to the timing sequence of one typical experiment, which is repeated over to gather statistics. The repetition rate is 13 ms, with one experimental run includes laser cooling and trapping atoms in a magneto-optics trap (MOT), then optically pumping the atoms to the initial state $\ket{2}=\ket{F=1,m_F=0}$ with or without a magnetic field ($B_z$) and decay beams for $\gamma_g$ and $\gamma_d$, and then storing and retrieving probe pulses by turning on and off the control fields. During the MOT cycle $^{87}Rb$ atoms are cooled and trapped in $\ket{F=2}$ hyperfine manifold. After trapping and further cooling the atoms with few ms of polarization gradient cooling (PGC) we start the cycle of storage of the probe pulse. Before it starts, we ensure  all the cooling and trapping beams as well as the MOT coil currents are turned off and the resulting fields are nulled. 

All this takes 10.1 ms: at 8.9 ms, the MOT coil currents are turned off to let enough time for ringing down of the trapping magnetic fields- at 9.8 ms, the frequency of the MOT beams is decreased by $\sim$80 MHz for efficiently pumping the atoms from $\ket{F=2}$ to $\ket{F=1}$- at 10.1 ms, the repumper field ($\ket{F=1}\rightarrow\ket{F'=2}$) is turned off and a near resonant optical beam ($\ket{F=2}\rightarrow\ket{F'=2}$) is turned on for 200 $\mu$s to optically pump the atoms to $\ket{F=1}$ ground state- all the beams are then turned off at 10.3 ms- in the dark, stray magnetic fields are cancelled with X,Y and Z bias coils- at 11.170 ms (at this instance the timing shifts to digital FPGA board with ns timing resolution from DAQ analog control), after shimming the magnetic fields, two orthogonal circular polarized light beams are turned on for 10 $\mu$s: the fields are near resonant from $\ket{F=1,m_F=-1}\rightarrow\ket{F'=0,m_{F'}=0}$ and $\ket{F=1,m_F=1}\rightarrow\ket{F'=0,m_{F'}=0}$. This completes the initialization of the atomic states, preparing the atoms in $\ket{F=1,m_F=0}$ ground state. The state preparation beams are turned off at 11.2 ms, and the storage cycle starts. In the storage cycle, a $\pi$-polarised probe pulse of 250 ns is turned on after 1.5 $\mu$s. A linearly polarized control field (composed of left and right circularly polarized fields) is turned on at the beginning of the storage cycle and is turned off at 1.6 $\mu$s in the storage cycle. The process of adiabatic turn off of the control field stores the probe pulse in the form of DSPs. After a controllable time, another linear polarized control field is turned on to read out the stored pulse. By changing the angle of polarization of the storing control field with respect to this retrieval field, we control the tone phase $\phi_c$.

To engineer the incoherent decays from ground state $\ket{F=1,m_F=-1}\rightarrow\ket{F=1,m_F=0}$ ($\gamma_g$) and $\ket{F=1,m_F=1}\rightarrow\ket{F=1,m_F=0}$ ($\gamma_d$), we use two red detuned ($4$ MHz) left and right circularly polarized fields that are deliberately phase mismatched by an angle with respect to the control fields. These counter propagating fields, inducing transition between the states $\ket{F=1,m_F=-1}\rightarrow\ket{F'=0,m_{F'}=0}$ and $\ket{F=1,m_F=1}\rightarrow\ket{F'=0,m_{F'}=0}$ are kept on for 3$\mu$s, from the beginning of the storage cycle. We set the total time of the storage experiment to 500 $\mu$s: in this timescale, we do not observe any atom loss due to radiation pressure of imbalanced control fields. 

Numerically reconstructed states and the corresponding Husimi-Q functions for each time instances are plotted in Fig.~\ref{fig:Motcycle}. Initially, $^{87}Rb$ atoms are cooled and trapped into $\ket{F=2}$ ground state in MOT cycle left with no atoms in $\ket{F=1}$ manifold at around 9.8 ms. After the atoms are optically pumped to ground state $\ket{F=1}$, the uniformly distributed atoms then correspond to uniform \textit{Q} over the entire surface of the sphere. At around 11.170 ms we turn on the polarizing beams (composed of two circularly polarized laser fields that are resonant from $\ket{F=1,m_F=\pm1}$ to $\ket{F'=0,m_{F'}=0}$) to prepare the atoms in the initial state $\ket{F=1,m_F=0}$. This state corresponds to equatorial limit cycle. The corresponding \textit{Q} function is distributed uniformly over all $\phi$ but centred around $\theta=\pi/2$. At 11.2 ms the main experimental sequence, corresponding to the storage of probe pulse, starts. During this sequence, we observe that for symmetric decay rates, when the DSPs interfere destructively, the \textit{Q} function remains distributed uniformly over $\phi$. However, the asymmetric decay rates induce synchronization, where the atomic, quantum phase $\phi$ locks to the tone phase of the classical, two photon drive. Finally, after the probe pulse is retrieved, the atomic state settles back to an equatorial limit cycle.

  \begin{figure*}
 	\includegraphics[width=\linewidth]{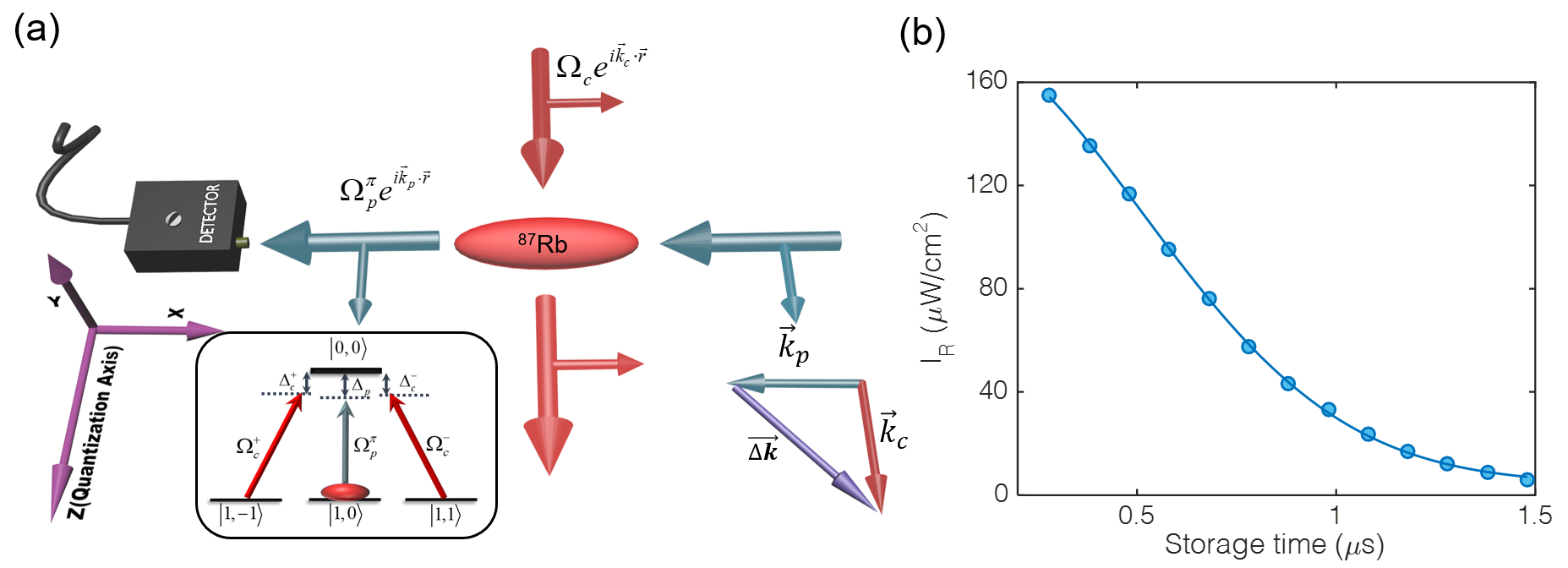}
 	\caption{(a) Simplified schematic of a typical storage experiment, where a linear polarized control field (red), comprising of two circularly polarised fields, is perpendicular to a $\pi$-polarized probe field. The resulting phase(momentum) mismatch $\Delta\vec{k}$ is shown for the fields. Inset: a schematic of four-level system is shown along with the control and probe field. (b) Experimental data of retrieved intensity as a function of storage time, with $\Delta_B=0$. The decaying retrieved pulse intensity is fitted to a Gaussian function ($ae^{-\tau^2/2\tau_{d}^2}$) of width $\tau_{d}\approx510$ ns. From the fitting, effective temperature of the atoms are estimated to be $121$ $\mu$K. Here, $I_{R}$: retrieved pulse intensity, $I_{i}$: input probe pulse intensity }
 	\label{fig:StorageTime} 
 \end{figure*}

\bigskip
\noindent
\textbf{(b) Doppler effect and storage in perpendicular configuration:} \label{DopplerEffect}

\bigskip
\noindent
For our experiment (Fig.~\ref{fig:StorageTime}(a), inset), we keep the control and probe fields perpendicular to each other (Fig.~\ref{fig:StorageTime}(a)). This perpendicular geometry results in excellent signal to noise ratio for detecting probe intensity in a direction that remains orthogonal to all other strong control and decay beams. However, such a geometry also results in a large mismatch of the two-photon momentum, causing severe Doppler decoherence. 
This time can be estimated from the time taken by an atom moving at average thermal velocity, to move over one period of the resulting wave-vector mismatch grating i.e.
\begin{equation}
\tau_{d}=\frac{\lambda_{s}/2}{v_{\text{th}}},
\label{eq:StorageTime}
\end{equation}
after substituting $\lambda_s$ and $v_{th}$ in equation~(\ref{eq:StorageTime}) we get
\begin{equation}
\tau_{d}=\frac{1}{2|\Delta\vec{{k}|}}\sqrt{\frac{\pi m}{2k_{B}T}}.
\end{equation}
From the measured $\tau_D$, we thereby estimate the average thermal velocity and the temperature of the atomic ensemble as:
\begin{eqnarray}
T=\frac{\pi m}{8|\Delta\vec{{k}}|^{2}k_{B}\tau_{d}^{2}},
	\label{eq:Dopplertime}
\end{eqnarray}
here $|\Delta\vec{{k}}|^{2}=|\vec{k}_{c}|^{2}+|\vec{k}_{p}|^{2}+2\vec{k}_{c}\cdot\vec{k}_{p}=|\vec{k}_{c}|^{2}+|\vec{k}_{p}|^{2}=2|\vec{k}_{p}|^{2}$, as direction of $\vec{k}_{p}$ and $\vec{k}_{c}$ is perpendicular to each other and $|\vec{k}_{p}|=|\vec{k}_{c}|$, $k_B$ is the Boltzmann's constant, and $\tau_d$ is the decoherence time due to the thermal velocity of atoms. In Fig.~\ref{fig:StorageTime}(b) the peak of the retrieved pulse intensity with respect to the storage time $\tau$ is plotted and fitted with a Gaussian function $ae^{-{\tau^{2}}/{2\tau_{d}^{2}}}$, as averaged over Maxwell velocity distribution yields Gaussian decay. From the fitting parameter, we extract $\tau_{d}\approx510$ ns. Plugging in the values of $\tau_d, k_B$ and $|\Delta\vec{k}|$ in equation~(\ref{eq:Dopplertime}), we estimate the temperature of the atomic cloud to be T$\sim121~\mu$K. 

\begin{flushleft}
\textbf{(c) Magnetic field calibration using interferometric measurement:}
\end{flushleft}

  \begin{figure*}
 	\includegraphics[width=\linewidth]{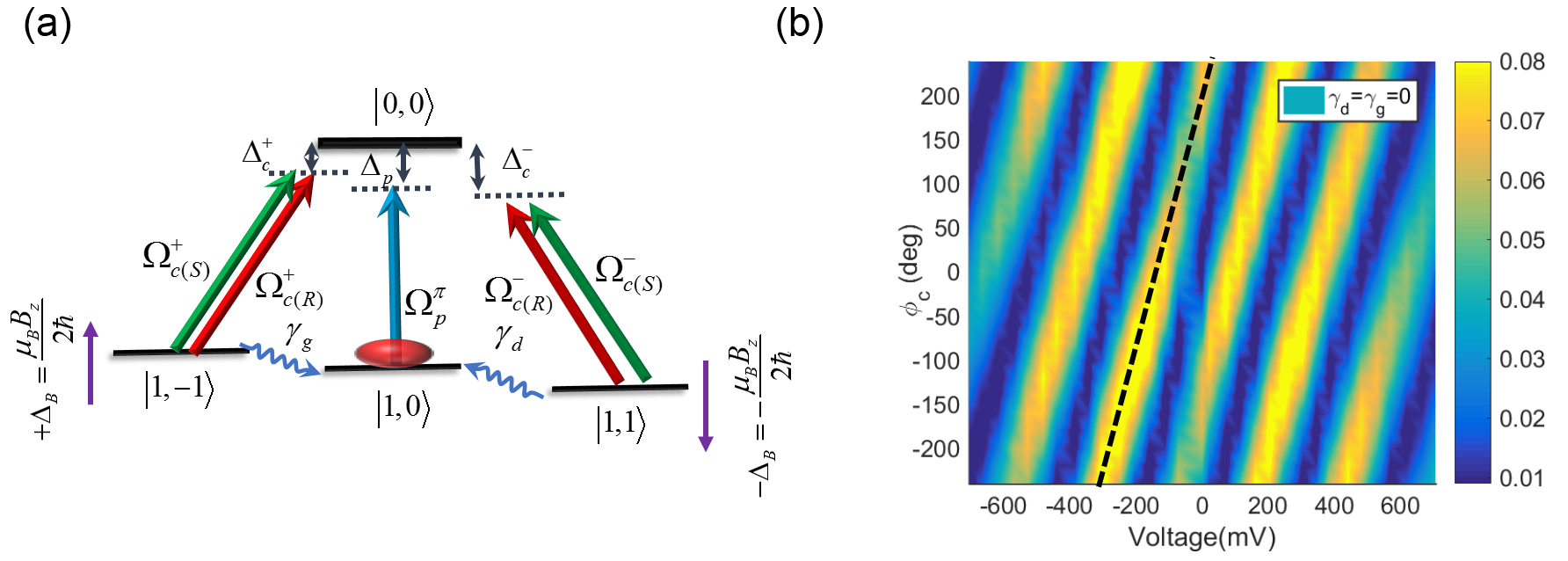}
 	\caption{(a) Atomic energy level scheme of the 4-level system. A $\pi$-polarized Probe $\Omega_{p}^\pi$ (blue) is stored with a linearly polarized control field $\Omega_{C(S)}^\text{lin}$ (green) composed of two orthogonal circularly polarized fields $\Omega_{c(S)}^+$ and $\Omega_{c(S)}^-e^{i\phi_S}$, and retrieved by another linearly polarized control field $\Omega_{C(R)}^\text{lin}$ (red) (composed of $ \Omega_{c(R)}^+$ and $\Omega_{c(R)}^-e^{i\phi_R}$). The dynamic phase ($2\Delta_B\tau$) between the two dark state polaritons which are stored in the medium is due to an applied magnetic field $B_z$ along the quantization axis. (b) Peak of the retrieved pulse intensity at retrieval time ($\tau = 600~\text{ns}$) is plotted with applied magnetic field (measured voltage due to the current though the coils) and control phase $\phi_c=\phi_S-\phi_R$, here we set $\phi_R=0$. We observe fringes due to interference between the two dark state polaritons. The slope of the black dotted line is used to calibrate the magnetic field and thereby, the dynamic phase ($2\Delta_B\tau$). 
 	}
 	\label{fig:MagneticFieldCal}
 \end{figure*}

\noindent
As discussed, the two stored DSPs interfere constructively or destructively depending upon the phase difference $\phi_c=\phi_S-\phi_R$. 
In presence of an applied magnetic field, the generated coherences $\rho_{-1,0}$ and $\rho_{0,1}$ between the states $\ket{F=1,m_F=-1}$ , $\ket{F=1,m_F=0}$ and $\ket{F=1,m_F=1}$ acquire additional dynamic phase $2\Delta_B\tau(2\pi)$ in a finite storage time ($\tau$). 
Accordingly, the intensity of the retrieved pulse depends on the total phase $\phi_c+2\Delta_B\tau(2\pi)$. 
For all our experiments, we keep $\phi_R=0$ i.e, $\phi_c=\phi_S$. 
At a fixed magnetic field, we observe interference fringes in time for the retrieved pulse amplitude which we interpret as due to the dynamic phase difference between the two generated DSPs (Fig.~\ref{fig:ExpSimulation}(c)). Fig.~\ref{fig:MagneticFieldCal}(b) shows the retrieved pulse peak with $\phi_S$ and measured voltage $V_m$ from the current monitor of the current controller circuit which is used to generate magnetic field along the quantization axis. We use these interference fringes (Fig.~\ref{fig:MagneticFieldCal}(b)) to calibrate the Zeeman shift $\Delta_B$ to the measured voltage $V_m$ at the magnetic field controller. To relate $V_m$ to $\Delta_B$, we measure the peak of the retrieved pulse amplitude along the black line in Fig.~\ref{fig:MagneticFieldCal}(b). The peak is observed to depend on the total relative phase $\phi_S+2\Delta_B\tau(2\pi)=0$ between the two stored coherences or DSPs. Here the measured voltage ($V_m$) correspond to the dynamic phase $2\Delta_B\tau(2\pi)$. Along the black locus, the total change in $\phi_c$ (y-axis) and measured voltage $V_m$ (x-axis) are $480^0$ and 364 mV, respectively. So the equivalent change in phase per mV is $\frac{d\phi_S}{dV_m}=0.023$ $\text{rad}/\text{mV}$. Accordingly, the acquired dynamic phase can be written as:
\begin{equation}
4\pi\Delta_B\tau=\frac{d\phi_S}{dV_m}V_m ~\text{rad}. 
\end{equation} 
\noindent
After rearranging the terms we find:
\begin{equation}
\Delta_B=\frac{0.023}{4\pi\tau}V_m ~ \text{MHz},
\end{equation}
\noindent
where $V_m$ is in mV.

\begin{flushleft}
\textbf{(d) Engineering decay channels:}
\end{flushleft}
 \begin{figure*}
 	\includegraphics[width=\linewidth]{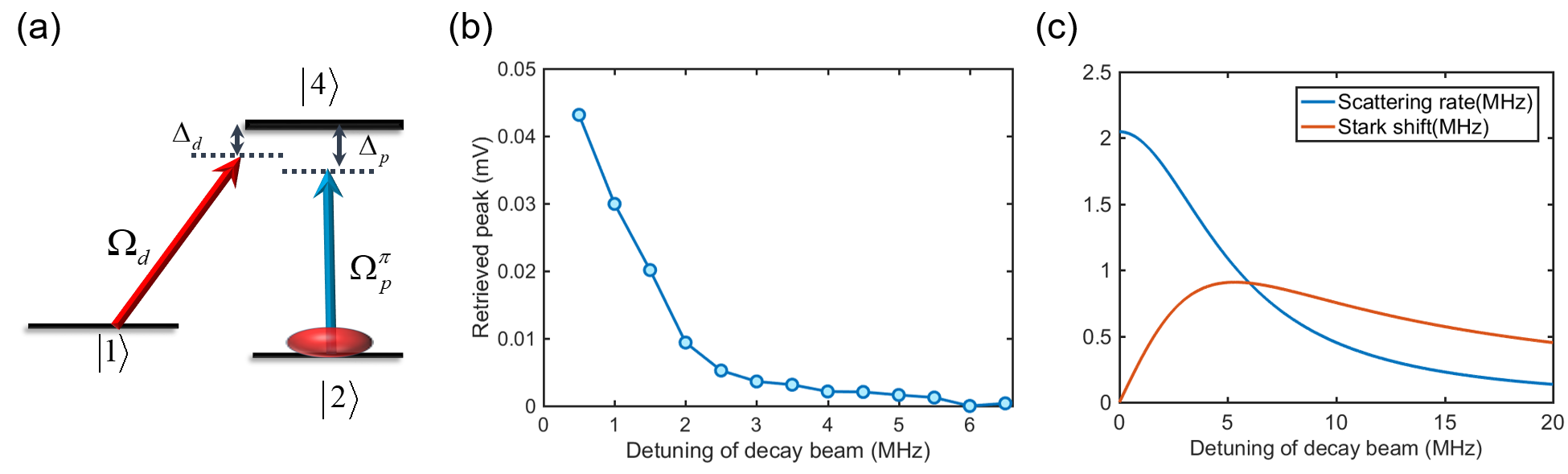}
 	\caption{(a) A schematic of the experiment to test induced coherence of the decay beams, with $\Omega_d$ as the decay beam along with a probe field $\Omega^\pi_p$. We mimic a storage and retrieval experiment by turning on and off the decay field. The resulting retrieved pulse is plotted in (b) as a function of the detuning $\Delta_d$ of the decay beam. From this plot, we choose $\Delta_d$= 4 MHz, where the retrieval, and thereby the induced coherence is less than 3\%. For a larger choice of $\Delta_d$, the decay rates $\gamma_g$ and $\gamma_d$ decreases significantly, as is shown the theoretical plots in (c). }
 	\label{fig:DecayBeamStorage} 
 \end{figure*}

 \begin{figure*}
 	\includegraphics[width=\linewidth]{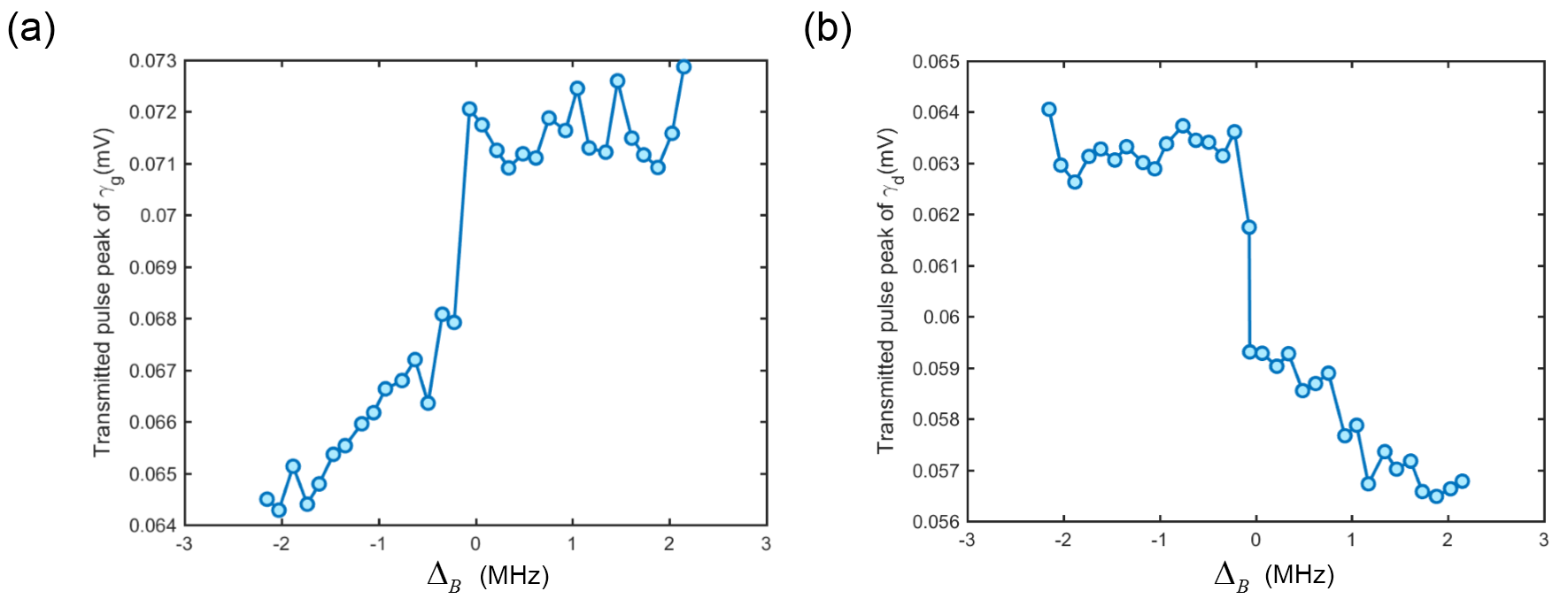}
 	\caption{(a) Transmission of decay beam $\gamma_{g}$ as a function of $\Delta_B$ which results in effective change in detuning with respect to excited state $\ket{F'=0,m_{F'}=0}$. (b) Same data but for $\gamma_{d}$. We use this data, to keep the detuning of the decay rates constant, by actively compensating shift due to an applied magnetic field. }
 	\label{fig:DecayBeamabs} 
 \end{figure*}

 \begin{figure*}
 	\includegraphics[width=\linewidth]{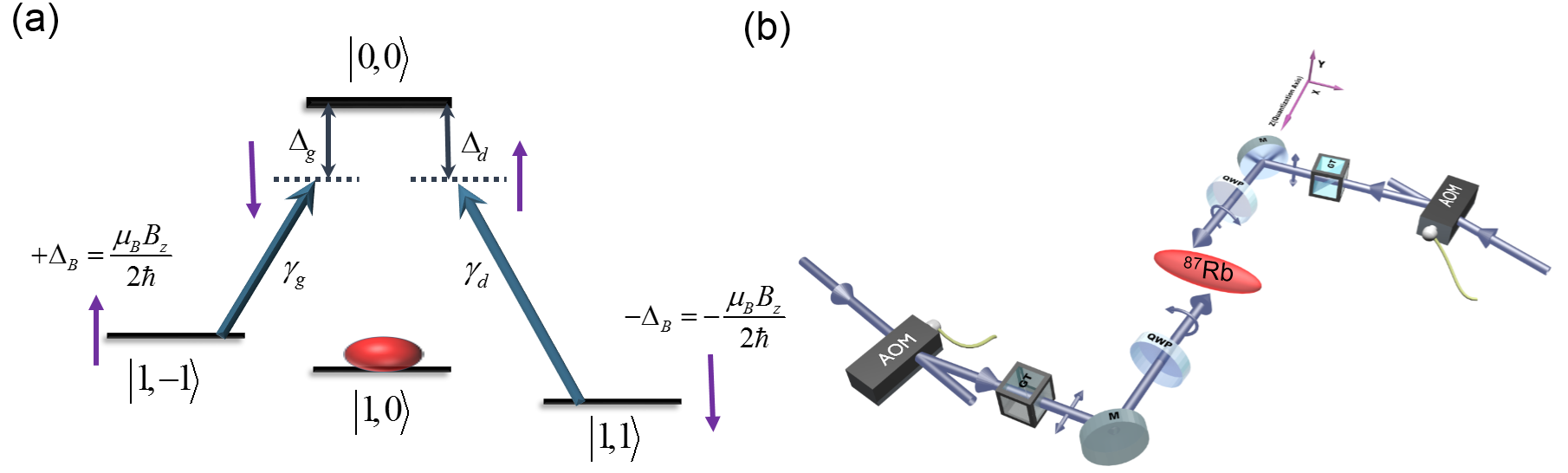}
 	\caption{(a) With increasing magnetic field in positive direction $\ket{1,-1}$ shifts upwards resulting into $\Delta_{g}<$ 4 MHz, and the state $\ket{1,1}$ shifts downwards resulting $\Delta_{d}>$  4 MHz. So, the detuning of decay beams $\gamma_{g}$ and $\gamma_{d}$ are kept fixed at 4 MHz by increasing and decreasing the detuning of the respective beams by tuning the AOMs drive frequency. (b) Schematic of experimental setup for decay beams. AOM: Acousto-Optic Modulator; GT: Glan-Thompson polarizing beam splitter; M: Mirror; QWP: Quarter-Wave Plate. }
 	\label{fig:DecayBeamCompansate} 
 \end{figure*}

\noindent
The decay beams (Fig. 1(b), main text) in the main text are used to create a stable equatorial limit cycle in a \textit{spin-1} system (Fig. 1, main text). These incoherent decays are generated from two lasers, 4 MHz red detuned from $\ket{F'=0}$, counter propagating and are left and right circularly polarized. Deliberate phase mismatch from the control and counter propagation ensures minimal generation of coherence~\cite{arif16_s} between $\ket{F=1,m_F=1}$ and $\ket{F=1,m_F=-1}$ states. 

However, in presence of probe, these fields can also independently generate ground state coherences $\rho_{-1,0}$ and $\rho_{0,1}$ in a EIT scenario. To avoid such a situation, we break the two-photon resonance condition for such processes by maintaining these beams red detuned from the probe by 4 MHz. To ensure this, we perform a typical storage experiment with $\pi$-polarized probe beam, and each of these circularly polarized decay beams (Fig.~\ref{fig:DecayBeamStorage}(a)). Here the states $\ket{1}, \ket{2}$ and $\ket{4}$ correspond to $\ket{F=1,m_F=1}$, $\ket{F=1,m_F=0}$, and $\ket{F'=0,m_{F'}=0}$, respectively. We keep the probe frequency resonant to the excited state and change the detuning of the decay beam. Fig.~\ref{fig:DecayBeamStorage}(b) shows the retrieved pulse peak with decay beam detuning. The detuning is increased to the red side of the excited state. The strength of the decay beam is $\Omega_d/\gamma=1.01$, where $\gamma$ is the excited state decay. The plot shows that after 3 MHz detuning, the retrieval efficiency has gone down significantly. The decay beams then essentially scatters atoms and introduce stark shifts of ground state levels. The scattering rates ($R_{sc}$) and stark shifts ($\Delta_{ss}$) due to this beam can be estimated from~\cite{Steck_s,Cohen98_s}: 
\begin{equation}
R_{sc}=\frac{\Gamma}{2}\frac{I/I_{sat}}{1+4\left(\delta/\Gamma\right)^2+I/I_{sat}},
\end{equation}
and 
 \begin{equation}
\Delta_{ss}=\frac{\delta}{2}\frac{I/I_{sat}}{1+4\left(\delta/\Gamma\right)^2+I/I_{sat}},
 \end{equation}
respectively. 
 
In Fig.~\ref{fig:DecayBeamStorage}(c) the scattering rates and stark shifts are plotted with detuning $\left(\Delta_d\right)$. We find a red detuning of 4 MHz  to be optimum choice, where the coherence falls down to less than few percent while the scattering rate still remains substantial with a fall of $50\%$, at this value of the detuning. Further increase of detuning leads to significant decrease in scattering rates. We keep the detuning of decay compensating beam at $4$  MHz throughout the experiment. 
 
 It can be noted that changes in magnetic field leads to ground state energy shifts ($\Delta_B$) which change the detuning of the decay beam from 4 MHz. Accordingly, scattering rate due to these beams also change. To keep the scattering rate constant, the detuning of these beams are adjusted to keep them at 4 MHz red detuned from the excited state. With positive applied magnetic field the detuning between $\ket{1,-1}$ and $\ket{0,0}$ reduces whereas detuning corresponding to $\ket{1,1}\longrightarrow\ket{0,0}$ enhances. To compensate for the change in detuning of decay beams, it is important to identify the exact polarization of two beams, and with the help of this information, we can adjust the detuning of beams accordingly. To identify in which direction the detuning must be compensated, we perform an experiment by observing the absorption of these decay beams with changing magnetic field by labelling them as $\gamma_{g}$ and $\gamma_{d}$. The transmitted peaks of $\gamma_{g}$ and $\gamma_{d}$ pulses are plotted with respect to detuning in Fig.~\ref{fig:DecayBeamabs}(a) and \ref{fig:DecayBeamabs}(b), respectively. It can be observed from Fig.~\ref{fig:DecayBeamabs}(a) that the absorption is increasing in the negative $\Delta_B$ side in case of $\gamma_{g}$, and it is identified as right circularly polarized light. With a similar argument, $\gamma_{d}$ is identified as left circularly polarized light (Fig.~\ref{fig:DecayBeamabs}(b)). As we change the detuning of the system from -2.15 MHz to 2.15 MHz, the effective detuning of $\gamma_{g}$ is fixed at 4 MHz by adjusting the detuning of the beam from $\left(4+2.15\right)=6.15 $ MHz to $\left(4-2.15\right)=1.85 $ MHz, and for $\gamma_{d}$ the change is opposite (Fig.~\ref{fig:DecayBeamCompansate}(a)). The detuning of the beams are adjusted by changing the AOM drive frequency. We derive these decay beams from a single laser which is locked using beat note lock technique~\cite{Schunemann99_s}. The output of the laser is divided into two paths and send through acousto-optic modulators (AOMs) for switching the intensities. The first order of the blue shifted beams from these AOMs are used as the decay beams (Fig.~\ref{fig:DecayBeamCompansate}(b)). These beams are resonant to the excited state $\ket{F'=0}$ when the AOMs are driven with 80 MHz RF sources. Here, we adjust the scattering rates ($\gamma_g,\gamma_d$) by controlling the frequencies and amplitude of RF signal which drive these AOMs.

\end{widetext}

\end{document}